\documentclass[nobib]{style}
\usepackage[english]{babel}
\usepackage[utf8]{inputenc}
\usepackage[normalem]{ulem}
\usepackage{graphicx}
\usepackage{framed}
\usepackage{amsmath}
\usepackage{amsthm}
\usepackage{amssymb}
\usepackage{amsfonts}
\usepackage{enumerate}
\usepackage{setspace} 
\usepackage{multirow} 
\usepackage{placeins} 
\usepackage{booktabs} 
\usepackage{enumitem} 
\usepackage{adjustbox} 

\usepackage[round,sectionbib]{natbib}
\usepackage{etoolbox}
\makeatletter
\renewcommand\hyper@natlinkbreak[2]{#1}
\makeatother

\newcommand{\indep}{\perp \!\!\! \perp}

\title{Functional Causal Inference with Time-to-Event Data}
\author{Xiyuan Gao$^1$, Jiayi Wang$^{2,*}$, Guanyu Hu$^3$, Jianguo Sun$^4$  \\  \vspace{5mm} \\
$^{1}$ \texttt{xgao@mail.missouri.edu} \\
$^{2,*}$ \texttt{jiayi.wang2@utdallas.edu} \\
$^{3}$ \texttt{guanyu.hu@missouri.edu} \\
$^{4}$ \texttt{sunj@missouri.edu} \\ \vspace{5mm} \\
$^{1,3,4}$ Department of Statistics, University of Missouri-Columbia \\
$^{2,*}$ Department of Mathematical Sciences, University of Texas at Dallas
}
\date{}     

\begin{document}
\maketitle
\begin{abstract}
Functional data is a powerful tool for capturing and analyzing complex patterns and relationships in a variety of fields, allowing for more precise modeling, visualization, and decision-making.
For example, in healthcare, functional data such as medical images can help doctors make more accurate diagnoses and develop more effective treatment plans.
However, understanding the causal relationships between functional predictors and time-to-event outcomes remains a challenge. 
To address this, we propose a functional causal framework including a functional accelerated failure time (FAFT) model and three causal approaches.
The regression adjustment approach is based on conditional FAFT with subsequent confounding marginalization, while the functional-inverse-probability-weighting approach is based on marginal FAFT with well-defined functional propensity scores.
The double robust approach combines the strengths of both methods and achieves a balance condition through the weighted residuals between imputed observations and regression adjustment outcomes.
Our approach can accurately estimate causality, predict outcomes, and is robust to different censoring rates. 
We demonstrate the power of our framework with simulations and real-world data from the Alzheimer's Disease Neuroimaging Initiative (ADNI) study. 
Our findings provide more precise subregions of the hippocampus that align with medical research, highlighting the power of this work for improving healthcare outcomes.
\end{abstract}
\begin{center}
\begin{minipage}{14truecm}
\end{minipage}
\end{center}
\vspace{3mm}

\section{Introduction}
Alzheimer's disease (AD) is a debilitating and fatal neurodegenerative disorder that affects millions of people worldwide, which has become the 6th leading cause of death among US adults \citep{CDC2022}, with increasing death rates and high costs of medical expenses \citep{AD2022}.
However, the pathogenesis of AD remains unclear, and there is currently no effective cure\citep{ashleigh2023role}. 
Early detection of mild cognitive impairment (MCI) and accurate assessment of progression from MCI to AD (MCI-AD) are crucial for the effective management of the disease \citep{potashman2023psychometric}.
Magnetic resonance imaging (MRI) has emerged as a valuable tool in the diagnosis and prognosis of AD, particularly in the analysis of the hippocampus which is a detectable brain atrophy susceptible to the impact of AD and frequently used as an MCI-AD progression biomarker. 
As a result, increasing attention has been given to exploring the use of MRI imaging data on the hippocampus to determine its relationship with MCI-AD and predict the time of conversion from MCI to AD \citep{warren2023functional}.

By treating the MRI on the hippocampus as a functional predictor, recent advancements including \citet{2018_Kong} and \citet{2021_Yang} investigated its conditional effects on the hazard ratio while assuming that the effects of other covariates are held constant in the functional linear Cox regression model, but left the causal relationship as an open question.
In an AD observational study such as Alzheimer's Disease Neuroimaging Initiative (ADNI), it is challenging to interpret the fitted models if the clinical measurements are confounded by other factors such as age, education, and ADAS-Cog score.
In order to explore the causal relationship between the hippocampus and MRI-AD, MRI imaging is regarded as a treatment throughout the paper to be in line with the causal inference language. 

The existing causal inference methods considering time-to-event data primarily focused on binary \citep{cao2023adjusting}, multi-level \citep{corder2022utilizing}, continuous \citep{cui2021instrumental}, and time-varying treatments \citep{yang2020semiparametric},
and have not addressed the complexities of functional treatments.
Unlike time-varying treatments that are measured sequentially at a sparse time grid in a longitudinal analysis, functional treatments measure the entire trajectory synchronously over the whole domain at one time point, without a dynamic component. 
They are featured by the high-dimension, continuous nature, and discrete observations, resulting in a non-sparse effect on outcomes.
By focusing on continuous responses, \citet{2021_Xiaoke} proposed the functional weights, which cannot be straightforwardly adapted to time-to-event data due to the existence of censoring.
Therefore, new methods are needed to account for unique features of functional treatment and determine its causal effect on time-to-event data.

The well-known Cox regression model has been widely investigated with a functional variable \citep{2018_Kong,2021_Yang}, whose clinical interpretation of hazard ratios (HR) is limited by the proportional hazard (PH) assumption.
However, even when the PH is plausible, the HR is not causally interpretable unless there is no treatment effect or an untestable and unrealistic assumption holds \citep{2022_Marti}.
As an alternative, the accelerated failure time (AFT) model was proposed and was able to provide a direct interpretation of the mean survival time \citep{1990_Ritov}.
It has been widely studied for a small number of predictors \citep{saikia2017review} but has not yet been generalized to functional data.
In order to fill the research gap, this paper aims to develop a novel functional causal framework that can effectively handle the functional treatment and provide a meaningful causal interpretation with time-to-event data.

The novel functional causal survival framework consists of the functional AFT (FAFT) model and three functional causal estimation approaches.
Distinguishing from previous research, this paper makes several new contributions.
The FAFT is proposed by incorporating functional linear regression into AFT for modeling the relationship between survival outcomes and functional treatment. 
We define the functional estimand as the coefficient of functional treatment in FAFT, representing the causal differences and variations over a domain. In fact this idea and our proposed approach are widely applicable to general time-to-event data with functional predictors.
To deal with the nonparametric functional treatment and corresponding coefficient, the functional principle component analysis (FPCA) is employed to achieve dimension reduction and avoid computation difficulty in regularization methods \citep{sang2022functional}.
To adjust the confounding effects, three causal inference approaches are proposed for time-to-event data, including the regression adjustment approach, the functional inverse-probability-of-weighting (FIPW) approach, and the double robust approach. 

The remainder of the paper is organized as follows.
In Section~\ref{sec:model_assumptions}, we first review the FPCA on a functional treatment, and then introduce the required assumptions and propose the FAFT model.
Details of three causal inference approaches are presented in Section~\ref{sec:causal_approaches}.
To facilitate computation, Section~\ref{sec:computation} develops the corresponding algorithms used for FAFT estimation, regression adjustment, FIPW, and double robust approach.
Extensive simulation studies are conducted in Section~\ref{sec:simulation} and an application on AD is included in Section~\ref{sec:real.data}.
In addition, the discussion can be found in Section~\ref{sec:discussion}. 
For ease of exposition, additional technical and numerical results are given in supplementary material.

\section{Model and assumptions}
\label{sec:model_assumptions}
\subsection{Notation and FPCA}
\label{sec:method_notaion_fpca}
The functional treatment $X(s)$ is defined over a compact set $\mathcal{S} \subseteq \mathbb{R}$, where $s$ denotes the continuous domain of $X(\cdot)$. However, in practice, $X(s)$ only can be observed on a discrete set of points within $\mathcal{S}$. 
The basic idea behind functional data analysis is to express discrete observed points in the form of a function, representing the entire measurements as a single observation. 
The functional linear regression considers a scalar response $Y$ and models its relationship with $X(\cdot)$ by using as follows:
\begin{equation}
Y  = \alpha + \int_\mathcal{S} \beta(s)X(s)\mathrm{~d}s + \epsilon,
\label{model:functional_linear_regression}
\end{equation}
where $\alpha$ is an unknown intercept, $\beta(\cdot)$ is an unknown coefficient with a functional form, and $\epsilon$ is an error term independent of $X(\cdot)$.
The variability of the functional coefficient $\beta(s)$ over $\mathcal{S}$ characterizes how the effect of a predictor variable changes smoothly over the entire domain.
It is also possible to interpret $\beta(\cdot)$ at specific points or regions.
In the AD example, $\beta(\cdot)$ allows us to study how different brain sub-regions impact the MCI-AD progression. 
By examining how the coefficient varies over the entire brain, we can gain insights into which sub-regions have the most significant impact and which are most vulnerable to progression.

Unlike scalar values, $X(\cdot)$ often exhibits complex structures such as high dimensionality, nonlinearity, and non-stationarity, making it difficult to visualize and analyze.
To overcome these challenges, this paper adopts FPCA due to its ability to represent functional data in the most parsimonious way \citep{jiao2021filtrated},
and then performs statistical analysis on the coefficients of these representations. 

Explicitly, the Karhunen-Loève theorem allows for the representation
$X(s)=\sum_{k=1}^{\infty} A_{k} \phi_{k}(s), \allowbreak s \in \mathcal{S},$
where $\{\phi_{k}(\cdot): 1 \leqslant k < \infty \}$ are eigenfunctions and $\{ A_{k} =\int_{\mathcal{S}} X(s) \phi_{k}(s) \mathrm{ d}s \}$ are functional principal component scores (FPCS), with the properties 
$E(A_{k})=0, \operatorname{Var}\left(A_{k}\right)=\lambda_{k}$, and $\operatorname{Cov}\left(A_{k_1}, A_{k_2}\right) = 0$ for any $k_1 \neq k_2$. 
The corresponding eigenvalues $\{\lambda_{k}: 1 \leqslant k < \infty \}$ satisfy that $\lambda_{k} \geqslant 0 $ in a decreasing order and $\sum_{k=1}^{\infty} \lambda_k < \infty$. 
Based on the span of $\phi_{k}(\cdot)$'s, the projection of $\beta(\cdot)$ corresponding to $X(\cdot)$ is also identifiable, as shown below:
$$
\int_\mathcal{S} \beta(s) X(s) \mathrm{~d}s = 
\int_\mathcal{S} \beta(s) \left( \sum_{k=1}^{\infty} A_{k} \phi_{k}(s) \right) \mathrm{~d}s = 
\sum_{k=1}^{\infty} A_{k} \left( \int_\mathcal{S} \beta(s) \phi_{k}(s) \mathrm{~d}s \right ).
$$
Let $\beta_k = \int_{\mathcal{S}} \beta(s)\phi_k(s)\mathrm{~d}s$, the expanded functional coefficient is
$\beta(s)=\sum_{k=1}^{\infty} \beta_k \phi_{k}(s), s \in \mathcal{S}.$

In practice, the number of FPCS used in modeling the data is usually truncated at some number $K$ due to computational constraints and to avoid overfitting. 
The selection of $K$ is typically based on the percentage of variation explained (PVE), which measures the proportion of the total variability in the data that is captured by the first $K$ FPCS.
In a sample of size $n$, the selected threshold is denoted as $K_n$ and calculated as PVE$(K_n) = \sum_{k=1}^{K_n} \lambda_k / \sum_{k=1}^{\infty} \lambda_k$. 
Therefore, $X(s) \approx \sum_{k=1}^{K_n} A_{k} \phi_{k}(s), s \in \mathcal{S},$ and the accuracy of such approximation has been proved to increase asymptotically as $n\to \infty$ \citep{wang2016functional}. 
Common choices of PVE$(K_n)$ range from $70\%$ to $99\%$, depending on the complexity of the data and the research question of interest \citep{2018_Kong}.

\subsection{Causality identification}
When observing a time-to-event response, take $T $ and $C$ as the failure time and censoring time, respectively. 
The observed event time is denoted as $\widetilde{T}=\min(T, C)$ and the corresponding censoring indicator is determined by $\delta = \operatorname{I}\allowbreak(T \leqslant C)$, which equals to 1 if observing a failure and 0 otherwise.
To remove the non-negativity restriction on event times, we define the response on its $\log$ scale, i.e., $Y = \log T$. 
In addition, a $p-$dimensional vector of observed covariates is represented by the vector $\mathbf{Z}$.

In accordance with the potential outcomes framework \citep{rubin1974estimating}, consider a potential treatment $x := x(s) \in L^2 = \{f: \int_{\mathcal{S}} f^2(s)\mathrm{~d}s < \infty\}$.
Let $Y(x)$ be the potential outcome and $\beta(\cdot)$ be the functional coefficient representing the causal effect curve.
Our objective is to model $\mathbb{E} \left[ Y(x) \right]$ and estimate $\beta(\cdot|x)$, in terms of the observed data
$\mathcal{O} = \Bigl \{ \left(X_i(s), \mathbf{Z}_i, \delta_i, \widetilde{T}_i\right) , \ s \in \mathcal{S},\allowdisplaybreaks  i=1,...,n \Bigr\}$.
To link the observed data to counterfactual data, the following assumptions are required to identify the causal effect.
\begin{itemize}[leftmargin=*,align=left]
    \item[\textbf{Assumption 1.} \emph{Ignorability }] 
    $ \quad X \indep Y(x)\ |\ \mathbf{Z} ,\ \forall x \in L^2.$
\end{itemize}
This is also known as conditional exchangeability, i.e., $\mathbb{E}\left[ Y(x)|X,Z\right] = \mathbb{E}\left[Y(x)|\mathbf{Z} \right]$. 
In a randomized clinical trial (RCT), it always hold since the treatment allocation mechanism provides no information on the counterfactual outcomes.
In an observational study, this is generally not true and is said to hold provided that the treatment is randomized within strata of the recorded covariates. Such condition is generally not verifiable empirically and their plausibility should be justified based on subject matter knowledge in practice.
In general, it holds if the study guarantees it by design (e.g., stratified randomized trial) or sufficient potential confounders have been collected.

The following two assumptions are associated with the propensity score for a functional treatment.
Since the conditional density function of a functional treatment does not exist, the rank-$K$ functional propensity score is defined as $\pi_{\mathbf{Z}}(\mathbf{A}) = P(\mathbf{A}=\mathbf{a}|\mathbf{Z})$ by using the selected FPCS vector $\mathbf{A} = \left \{ A_1, ..., A_K \right \}$ \citep{2021_Xiaoke}.
\begin{itemize}[leftmargin=*,align=left]
    \item[\textbf{Assumption 2.} \emph{Consistency }]
    $\  \sum_{k=1}^{K} A_{k} \phi_{k}(s) = \sum_{k=1}^{K} a_{k} \phi_{k}(s), s \in \mathcal{S} \Rightarrow Y = Y(x).$ 
    \item[\textbf{Assumption 3.} \emph{Positivity }] 
    $\quad \   0<\pi_{\mathbf{Z}}(\mathbf{A})<1 \text{ with probability } 1, \quad \forall \mathbf{a} \in \mathbb{R}^{K}.$
\end{itemize}
Assumption 2 states that potential outcomes are uniquely defined by an subject's own treatment level, no interference with others and no different versions of treatment. 
This ensures that the causal effect of $X(\cdot)$ on an subject is not influenced by external factors.
By implying that every subject has a positive chance of receiving any level of treatment, regardless of their covariates, Assumption 3 ensures that the treatment effect estimates are derived from a representative sample of the population, rather than a biased subgroup of subjects.
It is important to point out that how to address the positivity violation remains an open problem for continuous treatments  \citep{zhao2020propensity}, let alone functional treatments, where the issue of positivity violation is likely to be more severe.

For the censoring mechanism, we assume independent censoring between the actual survival time and the censoring time.
\begin{itemize}[leftmargin=*,align=left]
    \item[\textbf{Assumption 4.} \emph{Noninformative censoring }]
    $ \ C \indep Y(x) | \left(\mathbf{Z}, X \right) \Rightarrow  C \indep Y(X)| \left(\mathbf{Z}, X \right). $
\end{itemize}

Given Assumptions 1–4, based on the observed data, $\mathbb{E}[Y(x)]$ can be identified via
\begin{equation}
    \mathbb{E}[Y(x)] = \int_\mathbf{Z} \mathbb{E}\left[  Y(x) |  \mathbf{Z}=\mathbf{z} \right] \mathrm{~d} f_\mathbf{Z}(\mathbf{z}) = \int_\mathbf{Z} \mathbb{E}\left[  Y | X = x,  \mathbf{Z}=\mathbf{z} \right] \mathrm{~d} f_\mathbf{Z}(\mathbf{z}) ,
\label{eq:identification}
\end{equation}
where $f_\mathbf{Z}(\mathbf{z})$ is the joint density function of all covariates.

\subsection{FAFT model}
In RCTs, the causal effect estimation is straightforward, since RCTs may achieve sufficient control over confounding factors provided a good design, proper conduction, and enough enrollment.
We propose a FAFT model that incorporates the functional linear regression and classical accelerated failure time model.  
Specifically, for subject $i$, the FAFT has the form
\begin{equation}
Y_i = \log (T_i) = \alpha + \int_\mathcal{S} \beta(s)X_i(s)\mathrm{~d}s  + \epsilon_i.
\label{model:faft}
\end{equation}
For simplicity, assume that $X_i(s)$ can be fully observed on grid points $\{ s_{m} \in \mathcal{S}, 1 \leqslant m \leqslant M\}$ without any measurement error.
The generalization to situations with measurement error can be found in \citet{2018_Kong}.
With FPCA, the FAFT can be rewritten and approximated as
\begin{equation}
Y_i = \log T_i 
= \alpha + \sum_{k=1}^{\infty} A_{ik} \beta_k + \epsilon_i 
\approx \alpha + \sum_{k=1}^{K_n} A_{ik} \beta_k + \epsilon_i.
\label{model:faft_FPCA}
\end{equation}
Such approximation reduces the estimation of FAFT with an unknown curve to a classical AFT regression model with finite-dimensional predictors.
Estimation details will be discussed in Section~\ref{sec:computation_FAFT} and the resulting estimates include $\hat{\alpha}$ and $\{\hat{\beta}_1,...,\hat{\beta}_{K_n}\}$.
The resulted $\hat{\beta}(\cdot)$ is the estimated functional coefficient in the model (\ref{model:faft}) and intuitively represents the causality between $Y$ and $X(\cdot)$ in RCTs.

However, this is usually not true in observational studies due to the possible disparities in confounders on both $X(\cdot)$ and $Y$.
We propose three approaches to address confounding effects and denote the resulting estimators with extra subscripts, including 
(1) regression adjustment approach, $\Bigl\{ \hat{\alpha}_{\text{RegAdj}},  \hat{\beta}_{\text{RegAdj},1}, ..., \hat{\beta}_{\text{RegAdj}, K_n} \Bigr \}$ and $\hat{\beta}_{\text{RegAdj}}(\cdot)$; 
(2) FIPW approach, $\Bigl \{ \hat{\alpha}_{\text{FIPW}}, \allowbreak \hat{\beta}_{\text{FIPW},1}, ..., \hat{\beta}_{\text{FIPW}, K_n} \Bigr \}$ and $\hat{\beta}_{\text{FIPW}}(\cdot)$; 
and (3) double robust approach, $\Bigl \{ \hat{\alpha}_{\text{DR}}, \hat{\beta}_{\text{DR},1}, ..., \allowbreak \hat{\beta}_{\text{DR}, K_n} \Bigr \}$ and $\hat{\beta}_{\text{DR}}(\cdot)$.

\section{Causal inference approaches}
\label{sec:causal_approaches}
\subsection{Regression adjustment approach}
\label{sec:method_regadj}
In order to marginalize the confounding effects, the regression adjustment approach provides consistent estimates of contrasts (e.g. differences, ratios) by regressing $Y_i$ conditional on $X_i$ and $\mathbf{Z}_i$ followed by adjusting confounding effects.
By considering a full FAFT,
\begin{equation}
Y_i = \log T_i = \alpha + \int_\mathcal{S} \beta(s)X_i(s)\mathrm{~d}s  + \boldsymbol{\gamma}^\top \mathbf{Z}_i + \epsilon_i  \approx \alpha + \sum_{k=1}^{K_n} A_{ik} \beta_k + \boldsymbol{\gamma}^\top \mathbf{Z}_i + \epsilon_i.
\label{model:faft_FPCA_full}
\end{equation}
the association between $X(\cdot)$ and $Y$ conditioning on $\mathbf{Z}$ is revealed.
This conditional mean survival outcome is the conditional expectation $\mathbb{E} \left[  Y | X=x, \mathbf{Z}=\mathbf{z} \right]$ in equation (\ref{eq:identification}). 

Suppose the fitted FAFT model gives parameter estimates $(\hat{\alpha}, \hat{\beta}_1, ..., \hat{\beta}_{K_n}, \hat{\gamma}_1, ..., \hat{\gamma}_p)$, the confounding adjustment proceeds in the following two steps.
For each subject $i$, construct new responses adjusted by empirical mean of all confounders, 
\begin{equation}
    \widehat{Y}_{i} = \frac{1}{n} \sum_{j=1}^n \left( \hat{\alpha} + \int_\mathcal{S} \beta(s)X_i(s)\mathrm{~d}s + \hat{\boldsymbol{\gamma}}^\top \mathbf{Z}_j \right) \approx \hat{\alpha}  + \sum_{k=1}^{K_n} A_{ik} \hat{\beta}_k + \frac{1}{n} \sum_{j=1}^n \hat{\boldsymbol{\gamma}}^\top \mathbf{Z}_j.
\label{eq:regadj.new.y}
\end{equation}
Based on $\left(X_i, \widehat{Y}_{i},i=1,...,n\right)$, refit model (\ref{model:faft_FPCA}) and the estimates $\Bigl\{ \hat{\alpha}_{\text{RegAdj}},  \hat{\beta}_{\text{RegAdj},1}, ..., \allowbreak \hat{\beta}_{\text{RegAdj},K_n} \Bigr \}$ can be used to reconstruct the estimated functional coefficient, via $\hat{\beta}(\cdot)_\text{RegAdj}=\sum_{k=1}^{K_n} \hat{\beta}_{\text{RegAdj},k} \phi_{k}(\cdot)$.

\subsection{FIPW approach}
\label{sec:method_fipw}
Instead of modeling survival outcomes conditionally as discussed in Section~\ref{sec:method_regadj}, the causal estimator can also be identified by direct marginal modeling. 
By defining the true weights to be $\mathrm{w} = f(\mathbf{Z}) / f(\mathbf{Z}|X)$,  the FIPW approach  created a weighted pseudo sample and the marginal mean of the potential outcome within it is equal to the adjusted conditional mean of the actually observed population. 
\begin{itemize}[leftmargin=*,align=left]
    \item[\textbf{Proposition 1.}] 
    Under Assumptions 1-3, if the sample is re-weighted by using the weights defined as $\mathrm{w} = f(\mathbf{Z}) / f(\mathbf{Z}|X)$, then in the created pseudo-sample, 
    \begin{equation}
    \mathbb{E} \Bigl[ \mathrm{w} Y | X(s)=x(s) \Bigr] = \mathbb{E}[ Y(x)]\Bigr].
    \label{eq:fipw_weights_requirements}
    \end{equation}
\end{itemize}
The left side of the equation is the expectation over the weighted pseudo-sample and is the logic behind FIPW, while the right side is the expectation after marginalization over $\mathbf{Z}$ and is the rationale behind the regression adjustment.
Proof details are included in Section 1.1 of the supplementary material.

Unlike categorical treatments whose propensity scores are defined in terms of conditional densities and are usually estimated directly, directly estimating the weights via the definition of  $w$ can be very challenging as the (conditional) densities ($f(Z)$ and $f(Z \mid X$)) are difficult to estimate and $Z$ is often multi-dimensional. 
Inspired by \citet{2021_Xiaoke}, the functional propensity score (FPS) and corresponding weight are defined as
$$
s_{i} = f_{\mathbf{A} \mid \mathbf{Z}}(\mathbf{a_i} \mid \mathbf{z_i}),\quad  w_{i}= \frac{f_{\mathbf{A}}\left(\mathbf{A}_{i}\right)}{s_{i}} = \frac{f_{\mathbf{A}}\left(\mathbf{A}_{i}\right)}{f_{\mathbf{A} \mid \mathbf{Z}}(\mathbf{a_i} \mid \mathbf{z_i})},
$$
where $\mathbf{A} = (A_{1}, ..., A_{K^*})$ is the selected FPCS vector from FPCA and $K^*$ can be different from $K$.
$f_{\mathbf{A} \mid \mathbf{Z}}(\cdot)$ and $f_{\mathbf{A}}(\cdot)$ are the conditional and marginal probability densities of $\mathbf{A}$. 
This way to calculate weights is the same as the way to define $\mathrm{w}$, as derivation shown in Section 1.2 of the supplementary material.
\begin{itemize}[leftmargin=*,align=left]
    \item[\textbf{Proposition 2.}] 
    If $X(s) = \sum_{k=1}^{K^*} A_i \phi_k(s), s \in \mathcal{S}$, then 
    $w_{i} = \mathrm{w}_{i}$ for subject $i$.
\end{itemize}

For computation convenience, the standardized FPS $s_{i}^{*}$ and standardized weights $w_{i}^{*}$ are usually utilized.
Define
$s_{i}^{*} = f_{\mathbf{A}^{*} \mid \mathbf{Z}^{*}}(\mathbf{a_i}^{*} \mid \mathbf{z_i}^{*})$ and $ \  w_{i}^{*} = f_{\mathbf{A}^{*}}\left(\mathbf{A}_{i}^{*}\right)/s_{i}^{*}$
with standardized covariates $\mathbf{Z}_i^{*}=\boldsymbol{\Gamma}_{\mathbf{Z}}^{-1 / 2} \mathbf{Z}_i, \boldsymbol{\Gamma}_{\mathbf{Z}} = E\left(\mathbf{Z Z}^{\top}\right)$ and standardized FPCS $\mathbf{A}_{i}^{*}=\left(A_{i 1}^{*}, \ldots, A_{i K_n^*}^{*}\right)^{\top}, A_{ik_n}^{*}=\lambda_{k}^{-1 / 2} A_{ik_n}, \ k=1, \ldots, K_n^*$.
Apparently, the standardized FPS automatically satisfies the positivity and consistency assumption.
To satisfy the conditional exchangeability assumption, the weight vector $\boldsymbol{w}^{*}$ should be able to achieve the covariate balance condition in the sense that minimizing the weighted correlation between $\mathbf{A}^{*} $ and $\mathbf{Z}^{*}$, i.e., $\mathbb{E}\left(\boldsymbol{w}^{*} \mathbf{A}^{*} \mathbf{Z}^{*\top}\right) =0$.
The optimization details will be discussed in Section~\ref{sec:computation_FAFT}.
With the estimated weights $\boldsymbol{\hat{w}}$, FAFT regression is then performed on $X_i(\cdot)$ the weighted survival outcome $\hat{w}_i Y_{i}^{\text{Imputed}}$, where $Y_{i}^{\text{Imputed}}$ is obtained based on equation (\ref{faft:pseudo_outcome_all}), resulting in $\hat{\beta}_{\text{FIPW}}(\cdot)$.

\subsection{Double robust approach}
\label{sec:dr}
The regression adjustment approach discussed in Section~\ref{sec:method_regadj} requires the correct regression model specification, while the FIPW approach proposed in Section~\ref{sec:method_fipw} is unstable when outliers exist.  
Therefore, the double robust approach is proposed and the advantage of being less sensitive to model misspecifications and data outliers, and attaining faster rates of convergence when both models are consistently estimated \citep{van2003unified}. 
After the regression adjustment, for subject-$i$, let $Y_{i}^{\text{Imputed}}$ be the imputed outcome obtained during the estimation process of the full AFT and $\hat{Y}_{\text{RegAdj},i}$ the fitted causal outcome.
We construct a new adjusted pseudo outcome as follows,
\begin{equation}
\widetilde{Y}_i = \hat{Y}_{\text{RegAdj},i} + \hat{w}_i \left( Y_{i}^{\text{Imputed}} - \hat{Y}_{\text{RegAdj},i} \right),
\label{eq:doublerobust.new.outcome}
\end{equation}
By applying $\left\{ \widetilde{Y}_1, ..., \widetilde{Y}_n \right \}$ on model (\ref{model:faft_FPCA}), we can obtain the $\hat{\beta}_{\text{DR}}(\cdot)$.

\section{Implementations of the approaches}
\label{sec:computation}
The objective of Sections~\ref{sec:model_assumptions} and~\ref{sec:causal_approaches} is to present a comprehensive framework for identifying the causal relationship between a functional treatment and a survival outcome. 
In this section, we will provide in-depth information on the estimation process and summarizes the algorithms utilized within this functional causal survival framework.

\subsection{FAFT estimation}
\label{sec:computation_FAFT}
With simpler notation, let $\boldsymbol{\theta}$ represent all unknown parameters and $\mathbf{D}_i$ predictors, respectively. 
For example, $\boldsymbol{\theta} = (\alpha, \beta_1, ..., \beta_k)^\top$ in FAFT (\ref{model:faft_FPCA}) and $\boldsymbol{\theta} = (\alpha, \beta_1, ..., \beta_k, \gamma_1, ..., \gamma_p)^\top$ in full FAFT (\ref{model:faft_FPCA_full}).
There, we write models as $Y_i = \boldsymbol{\theta}^\top  \mathbf{D}_i + \epsilon_i$.

To fit the FAFT, we generalize the idea of the least squares method due to its stability for a large number of predictors \citep{jin2006least}.
To deal with the right censoring, each response $Y_i$ is imputed by its conditional expectation,
\begin{equation}
Y_{i}^{\text{Imputed}}(\boldsymbol{\theta}) = \delta_{i}Y_{i} + \left(1-\delta_{i}\right)\mathbb{E}_{\boldsymbol{\theta}}\left[Y_{i} \mid Y_{i} \geqslant \log C_{i}\right],
\label{faft:pseudo_outcome_all}
\end{equation}
where the expectation is evaluated based on the Kaplan-Meier estimator $\hat{F}(e)$. 
Let $r_i(\boldsymbol{\theta}) = \widetilde{T}_i \ - \  \mathbb{E}_{\boldsymbol{\theta}} \left[ Y | X_i, \mathbf{Z}_i \right]$, then 
$\hat{F}(e)=1-\prod_{\{i: \mathbf{r}(i) \leqslant e\}}\left(\frac{n-i}{n-i+1}\right)^{\delta_{i}},$
where $\mathbf{r}(i)$ are the ordered $r_i(\boldsymbol{\theta})$'s.
Therefore the expectation in (\ref{faft:pseudo_outcome_all}) is estimated as 
\begin{equation}
\hat{\mathbb{E}}\left(Y_{i} \mid Y_{i} \geqslant \log C_{i}\right) =  \mathbb{E}_{\boldsymbol{\theta}} \left[ Y | X_i, \mathbf{Z}_i \right] +  \hat{\mathbb{E}}\left[e_{i} \mid e_{i} \geqslant r_i(\boldsymbol{\theta})\right],
\label{faft:pseudo_outcome_censor}
\end{equation}
where $\hat{\mathbb{E}}\left[e_{i} \mid e_{i} \geqslant r_i(\boldsymbol{\theta})\right] = \int_{r_i(\boldsymbol{\theta})}^{\infty} \frac{s}{1-\hat{F}\left(r_i(\boldsymbol{\theta})\right)}\mathrm{~d} \hat{F}(s)$.
Given initial values $\hat{\boldsymbol{\theta}}^{(0)}$, the least square estimator is the solution of the estimating equation 
$ U_n(\boldsymbol{\theta}, \hat{\boldsymbol{\theta}}^{(0)})=\sum_{i=1}^n\left(\mathbf{D}_i-\bar{\mathbf{D}}\right)^{\top}\Bigl(Y_{i}^{*}(\hat{\boldsymbol{\theta}}^{(0)}) - \allowbreak \boldsymbol{\theta}^\top  \mathbf{D}_i\Bigr) \allowbreak =0. $
The estimation procedure can be proceeded iteratively by solving 
$\hat{\boldsymbol{\theta}}^{(m)}_n =  L_n\Bigl( \hat{\boldsymbol{\theta}}^{(m-1)}_n \Bigr), \allowbreak \\ m \geqslant 1$, and  
\begin{equation}
L_n(\boldsymbol{\theta})=\left[\sum_{i=1}^n\left(\mathbf{D}_i-\bar{\mathbf{D}}\right)^{\top}\left(\mathbf{D}_i-\bar{\mathbf{D}}\right)\right]^{-1}\left[\sum_{i=1}^n\left(\mathbf{D}_i-\bar{\mathbf{D}}\right)^{\top}\left(Y_{i}^{\text{Imputed}}(\boldsymbol{\theta}) - \bar{Y}_{i}^{\text{Imputed}}(\boldsymbol{\theta})\right) \right], 
\label{algorithm_aft_update}
\end{equation}
where $\bar{Y}_{i}^{\text{Imputed}}(\boldsymbol{\theta}) = \frac{\sum_{i=1}^n Y_{i}^{\text{Imputed}}(\boldsymbol{\theta})}{n}$.
The whole estimation procedure is briefly summarized in Algorithm 1 in the supplementary material.

In practical implementation, when the initial estimator is consistent and asymptotically normal such as the induced smoothing Gehan estimator, $\hat{\boldsymbol{\theta}}$ is also consistent and asymptotically normal \citep{jin2006least}. 
The multiplier resampling can be applied to approximate the variance of the resulting estimator.

\subsection{Causal effect estimation}
\subsubsection{Regression adjustment approach}
\label{sec:computation_regression_adjustment}
As details shown in Section~\ref{sec:method_regadj}, after fitting the full outcome regression FAFT model, adjusting the confounding effects takes two more steps: firstly reconstruct adjusted responses and then refit the FAFT.
Algorithm 2 in the supplementary material summarizes the whole procedure.

\subsubsection{FIPW approach}
Ideally, the functional weights are expected to achieve the balance condition for each observed subject, i.e.,
\begin{equation}
\mathbb{E}\left(w_i^{*} \mathbf{A}_{i}^{*} \mathbf{Z}_{i}^{*\top}\right) =0, \quad i=1,...,n.
\label{eq:ipw.balance.condition}
\end{equation}
The optimization process involves the selection of a tuning parameter, denoted as $\rho$, which represents the level of tolerance for unbalancing in order to resolve the non-convexity issue arising in equation (\ref{eq:ipw.balance.condition}).
The recommended value is $\rho_0=0.1/N$.
For further details, refer to \citet{2021_Xiaoke}. 
After getting $\hat{w}_{i}^{*}$ 's, according to equation (\ref{eq:fipw_weights_requirements}), a weighted pseudo-sample, which is created by $\boldsymbol{w} \circ \mathbf{Y} = \Bigl\{ w_1Y_{1}^{\text{Imputed}}, ..., w_nY_{n}^{\text{Imputed}} \Bigr\}$ with $Y_{i}^{\text{Imputed}}$ being calculated from equation (\ref{faft:pseudo_outcome_all}), and will be used to fit FAFT.
Algorithm 3 in the supplementary material is a summary of causal effect estimation.

\subsubsection{Double robust approach}
Since the double robust approach is a combination of regression adjustment and FIPW, the corresponding algorithm also involves the implementation of Algorithm 2 and Algorithm 3. The details are shown in Algorithm 4 in the supplementary material.

\section{A simulation study} 
\label{sec:simulation}
To evaluate the finite-sample performance of the proposed methods, we conduct simulation studies under two different scenarios, representing light and strong effects of confounding variables. 
In both scenarios, we consider three different censoring rates.

For the $i$-th subject, the functional treatment $X_i(\cdot)$ is given by 
$X_i(s) = \sum_{k=1}^{K=6} A_{ik} \phi_{k}(s), s \in [0,1].$
The six eigenfunctions are defined as $\phi_{2k-1}(s) = \sin{(2\pi ks)}, \phi_{2k}(s) = \cos{(2\pi ks)}, k=1,2,3$.
The corresponding FPCS are assumed to be $A_{i 1}=\sqrt{16} W_{i 1}, A_{i 2}=\sqrt{12} W_{i 2}, A_{i 3}=\sqrt{8} W_{i 3}, A_{i 4}=\sqrt{4} W_{i 4}, A_{i 5}=W_{i 5}$, and $A_{i 6}=W_{i 6}/\sqrt{2}$ , where  $W_{i 1}, ..., W_{i 6}$ are simulated from a multivariate normal distribution with mean $\mathbf{0}$ and covariance matrix $\text{diag}(\sqrt{K}, \dots, \sqrt{K})$ with $K=6$.
Suppose that a three-dimensional covariate vector $\mathbf{Z}_i = \left(Z_{i1},Z_{i2},Z_{i3}\right)^{\top}$ can be observed from each subject, simulating $\mathbf{Z}_i$ as $Z_{i1}=W_{i1}+e_{i1}, Z_{i2} = 0.2 W_{i2} + e_{i2}$, and $Z_{i3} = 0.2 W_{i3} + e_{i3}$ where $e_{i1} \sim \operatorname{N}(0,0.5)$ and $e_{i2}, e_{i3} \overset{\text{i.i.d}}{\sim}\operatorname{N}(0,1)$.
The log-transformed failure times $Y_i = \log T_i$ for all subjects are independently generated under two distinct scenarios: 
\begin{itemize}[leftmargin=1cm,align=left]
    \item Scenario 1: $Y_i = \log T_i = 1 + \int_0^1 \beta_0(s)X_i(s)\mathrm{~d}s  + 2 Z_{i 1} +  e_i,\ e_i \sim \operatorname{N}(0,0.5)$;
    \item Scenario 2: $Y_i = \log T_i = 1 + \int_0^1 \beta_0(s)X_i(s)\mathrm{~d}s  + 2 Z_{i 1} + 2 Z_{i 1}^2 \times A_{i 1} + e_i,\ e_i \sim \operatorname{N}(0,0.5)$;
\end{itemize}
in which the true functional coefficient is assume to be $\beta_0(s)= 2\sin (2 \pi s) +  \cos (2 \pi s)+ \sin (4 \pi s)/2 + \cos (4 \pi s) / 2, s \in[0,1]$, representing the true causal effect of $X(\cdot)$.
The survival times are then transformed back to the original scale. 
The right censoring is introduced by independently generating $C_i$ from $\text{Uniform}(a, b)$ with different $a$ and $b$ being considered to determine different right censoring rates. 
The observed survival outcome $\widetilde{T}_i$ is decided by $\min \{T_i, C_i\}$ and the censoring indicator is $\delta_i = I (T_i \leqslant C_i)$.
When the sample size of the study is $N$, the observed data are
$ \mathcal{O} = \Big\{ \widetilde{T}_i, \delta_i, X_i(s), Z_{i 1}, Z_{i2}, Z_{i3}, \quad i=1,...,N \Big \}. $

The analysis goal is to get the estimated causal effect $\hat{\beta}(\cdot)$ using $\mathcal{O}$.
The naive way fits the model~(\ref{model:faft_FPCA}) directly, assuming no confounding effect.
However, in both scenarios considered here, among the three generated covariates, $Z_{i1}$ is directly related to both $X_i(\cdot)$ and $Y_i$, while $Z_{i2}$ and $Z_{i3}$ are only directly related to $X_i(\cdot)$ but not $Y_i$.
Such existence of confounding variables may bring bias to the estimates of the treatment effect if not appropriately accounted for. 
Especially in scenario 2, the non-linear relationship between $Z_{i1}$ and $Y_i$ may pose additional challenges for modeling and estimation.
We will apply the  proposed methods to adjust confoundings and compare them with 
the naive approach.

Each scenario considered three levels of right censoring (20\%, 40\%, and 60\%) and two sample sizes ($N = 200$ and $400$), and each setup is repeated 500 times.
The selection criterion for both $K_n$ and $K_n^*$ is set to be PVE=95\%.

\subsection{Evaluation measures}
To assess the accuracy and validity of each estimator and provide insights into their strengths and weaknesses in handling linear and non-linear confounding effects, as well as right censoring, we consider the following measures.

\begin{itemize}[leftmargin=1cm,align=left]
    \item Define the relative mean square error as $\operatorname{RMSE} = \Bigl( \int_\mathcal{S} (\hat{\beta}(s) - \beta_0(s))^2\mathrm{~d}s \Bigr)  / \allowbreak  \Bigl( \int_\mathcal{S} \beta_0^2(s) \mathrm{~d}s \Bigr),$ where $\hat{\beta}(\cdot)$ is the estimated mean coefficient function over all simulation runs.
    A lower value indicates a more accurate estimate to the true functional causal effect.
   
   \item Define the integrated squared bias as $ \operatorname{ISB} = \left( \sum_\mathcal{S} (\hat{\beta}(s) - \beta_0(s))^2 \right) /  |\mathcal{S}| ,$ where $|\mathcal{S}|$ is the $L_2$ distance of the domain and $|\mathcal{S}|=1$ in our situation.  
   It characterizes the accuracy relative to the domain and a lower ISB indicates better accuracy.
   
   \item The averaged integrated squared error (AISE) is the mean of the integrated squared error (ISE) over all simulation runs.
   In the $r$-th simulation run, $\operatorname{ISE}_{\text{sim-r}} = \Bigl( \int_\mathcal{S} (\hat{\beta}_{\text{sim-r}}(s) - \beta_0(s))^2 \allowbreak \mathrm{~d}s \Bigr) / |\mathcal{S}|,$
   where $\hat{\beta}_{\text{sim-r}}(\cdot)$ is the estimated functional coefficient.
   The AISE illustrates the accuracy over all simulation runs and its standard error (SE) measures the estimation variation.
   By contrast, the median ISE (MISE) provides an additional summary of the accuracy, which is less sensitive to outliers than AISE.
\end{itemize}

To examine the prediction performance, the 80\%\ / 20\% splitting rule was applied on each generated dataset for training and testing.
In the $r$-th simulation run, the root mean squared error is calculated as 
$$ \operatorname{Root-MSE}_{\text{sim-r}} = \left(   \sqrt{ \sum_{i=1}^n (\hat{Y}_{\text{sim-r},i,\text{pred}} - Y_{\text{sim-r},i,\text{causal}})^2 / n } \right ),$$ 
where $\hat{Y}_{\text{sim-r},i,\text{pred}}$ is the fitted value and $Y_{\text{sim-r},i,\text{causal}}$ is the generated true causal outcome for subject $i$.
This measure assesses the accuracy of the predicted outcome compared to the true causal outcome.
We report its mean and three quantiles ($q_{.25}, q_{.50}$, and $q_{.75}$) over all runs.


\subsection{Simulation results}
The proposed three causal approaches gave rise to five causal estimators, namely RegAdj, FIPW.para (FIPW with weights estimated parametrically), FIPW.np (FIPW with weights estimated nonparametrically), DR.para (DR with weights estimated parametrically), and DR.np (DR with weights estimated nonparametrically). 
The results presented below are based on $N=400$, with similar results for $N=200$ included in Section 3 of the supplementary material.

\subsubsection{Results of estimation}
\label{sec:simulation_results_estimation}
\autoref{table:simluation.beta} summarizes four evaluation measures that assess the precision of the estimated functional coefficient $\hat{\beta}(\cdot)$ under two scenarios, each with three levels of censoring rates.
In scenario 1, where the confounding effect is weak and the dependence relationship is linear, all proposed causal estimators exhibit smaller errors than the naive estimator. 
When it comes to scenario 2 where the confounding effect is stronger and the correlation is more complex, the naive approach introduces greater bias to causal estimation.
Instead, the proposed methods and algorithms successfully handle the situation.
The much smaller values in means of RMSE, AISE, MISE, and ISB indicate a substantial improvement in estimation accuracy, and  the significant reduction in SE of AISE demonstrates the stability of the proposed algorithms. 

\begin{figure}[]
\centering
\includegraphics[width=\textwidth]{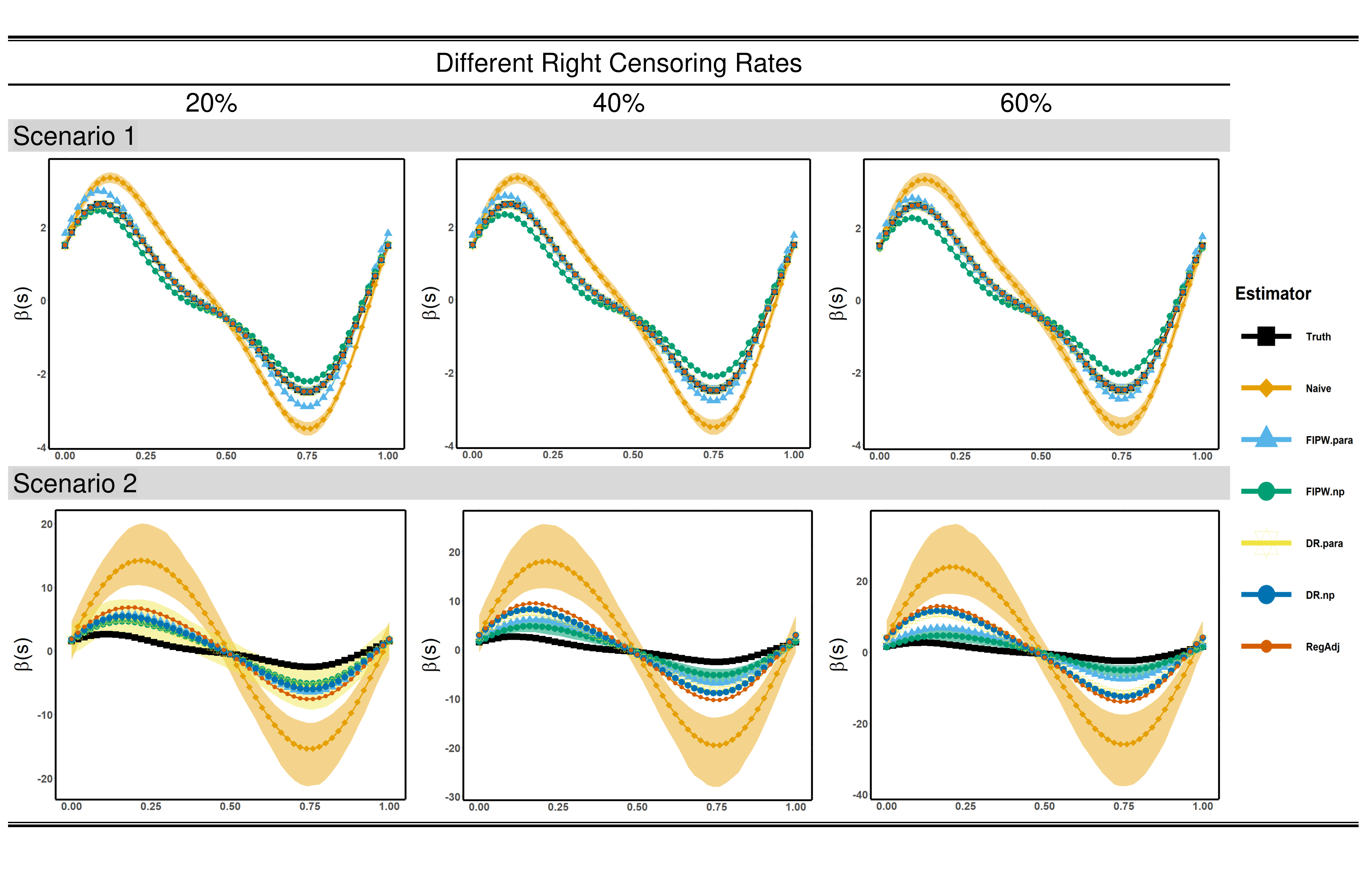}

\caption{Simulation results for evaluating the estimated functional causal effect curve under two scenarios with three censoring rates when sample size N=400. Different estimators are represented by line color and point shape with the shading area as the corresponding CI.}
\label{fig:simulation.beta.N400.plot}
\end{figure}

The estimated mean functional coefficients are displayed in \autoref{fig:simulation.beta.N400.plot}, with different estimators represented by variations in line colors and point shapes. 
The shaded area represents the confidence interval (CI) for $\hat{\beta}(\cdot)$ over all replications.
As shown in \autoref{fig:simulation.beta.N400.plot}, the naive estimator (the dark yellow line with dimond shapes) deviates away from the truth (the black line with squares) and has a 95\% CI that does not cover the truth in scenario 1, which is far worse in scenario 2.
By contrast, the five newly proposed estimators are shown to be close to $\beta_0(\cdot)$. 
In scenario 1, three estimators (DR.para in bright yellow, DR.np in dark blue, and RegAdj in orange) overlap with the truth, indicating less estimation bias than FIPW.para and FIPW.np. 
This is because the assumed model in the regression adjustment approach is similar to the true model used to generate our data. 
When such an assumption is heavily violated by the fact, which is the scenario 2, the FIPW approach (FIPW.para and FIPW.np) appears to have the advantage due to its ability to handle model misspecification. 
Another finding is that the FIPW approach yields a larger SE of AISE among five new estimators under scenario 1, while it becomes much smaller than others under scenario 2.
This difference is due to the possibility of extreme estimated weights, which add instability to the estimation process.
When the imbalance among confounding variables is weak (as in scenario 1), this instability contributes a larger proportion to the total variation, leading to a larger SE than others. 
In contrast, when the imbalance among confounding variables is stronger (as in scenario 2), the extreme estimated weights have less impact on the total variation, resulting in a smaller SE than others.

When looking at the DR approach, we can find that it performs between the regression adjustment and the FIPW approach, reflecting its "double robust" property.
This suggests that the DR approach outperforms the regression approach when the regression model is misspecified and is more stable when the estimated weights exist with extreme values.

\subsubsection{Results of prediction}
\autoref{table:simluation.prediction} includes in-sample and out-sample Root-MSE measures, displayed as the mean, 25\%, 50\%, and 75\% quantiles ($q_{.25}, q_{.50}$, and $q_{.75}$).
Compared to the naive estimator, the proposed approaches show better accuracy and efficiency in predicting causal outcomes, illustrated by a lower mean of Root-MSE and less variation among quantiles.
Even though all estimators decrease in performance with more complex confounding effects and higher censoring rates, the proposed methods always outperform the naive approach in a more robust way.
When comparing among five proposed estimators, once again, the double robust (DR.para and DR.np) and regression adjustment approaches show more accurate predictions in scenario 1, while the FIPW approach (FIPW.para and FIPW.np) performs better in scenario 2, as discussed in Section~\ref{sec:simulation_results_estimation}.

\subsubsection{Different right censoring rates}
Higher right censoring rates consistently lower estimation and prediction accuracy across scenarios and estimators due to information loss resulting from censoring. This creates uncertainty during imputation, leading to increased bias and variation in causal estimates and predictions.

As observed in \autoref{table:simluation.beta}, the SE of AISE for all estimators increases with increasing censoring rates under both scenarios.
While the mean of AISE remains relatively stable under scenario 1, it exhibits an obvious increasing trend under scenario 2, particularly there is a big jump in AISE and its SE in the transition from a 40\% to a 60\% censoring rate.
However, in comparison to the naive estimator, the proposed methods exhibit greater reliability and robustness.
In particular, the FIPW demonstrates superiority over the DR in scenario 2 as the censoring rate increases.
This is due to the increased bias that arises from utilizing the imputed outcomes of censored subjects as their true observations during the final step of estimating the DR estimators. 
This bias becomes more substantial and contributes to a decrease in estimation accuracy in the presence of a misspecified regression model in scenario 2.
A similar pattern can also be observed in \autoref{table:simluation.prediction} with increasing censoring rates.

\section{An application on AD}
\label{sec:real.data}
We implemented the proposed functional causal framework to analyze the data from the ADNI observational study, which comprises survival information about AD diagnosis, clinical measurements, and MRI records.
In contrast to previous studies, we aimed to identify the causal 
relationship between subregions within the hippocampus and the time of progression from MCI to AD, to contribute to a better understanding of the mechanisms underlying AD progression, and to provide insight into potential targets for early intervention.

For our purpose, 373 MCI subjects in ADNI-1 study along with their imaging and clinical measures are considered in the analysis. 
Among them, 161 MCI subjects progressed to AD during the study and the remaining 212 MCI subjects did not convert to AD prior to the study's end. 
Thus, the time of conversion from MCI to AD should be treated as time-to-event data with a censoring rate of 56.9\%.
The clinical characteristics of participants include age, education length, ADAS-Cog score, gender (0=male; 1=female), handedness (0=right; 1=left), marital status (0=single; 1=married), retirement (0=no; 1=yes), and apolipoprotein E (APOE) genetic covariates. 
The APOE is defined based on two single nucleotide polymorphisms (SNPs), rs429358 and rs7412, and produced a 3-allele haplotype, resulting in $\epsilon 2, \epsilon 3,$ and $\epsilon 4$ variants.
Descriptive statistics were summarized in \autoref{table:realdata.summary}, after categorizing the 373 MRI subjects into four groups based on their observed survival outcomes. 
The hippocampus image data was treated as a functional treatment and was represented as a matrix with dimension $(N,p) = (373, 30000)$ after being preprocessed \citep{2018_Kong}. 
Each row contained an subject's hippocampal radial distances of 30,000 surface points on the left and right hippocampus surfaces, which is a summary statistic of the hippocampal shape and size and defined as the distance between the medial core of the hippocampus and the corresponding vertex.

\begin{figure}[]
\centering
\includegraphics[width=\textwidth]{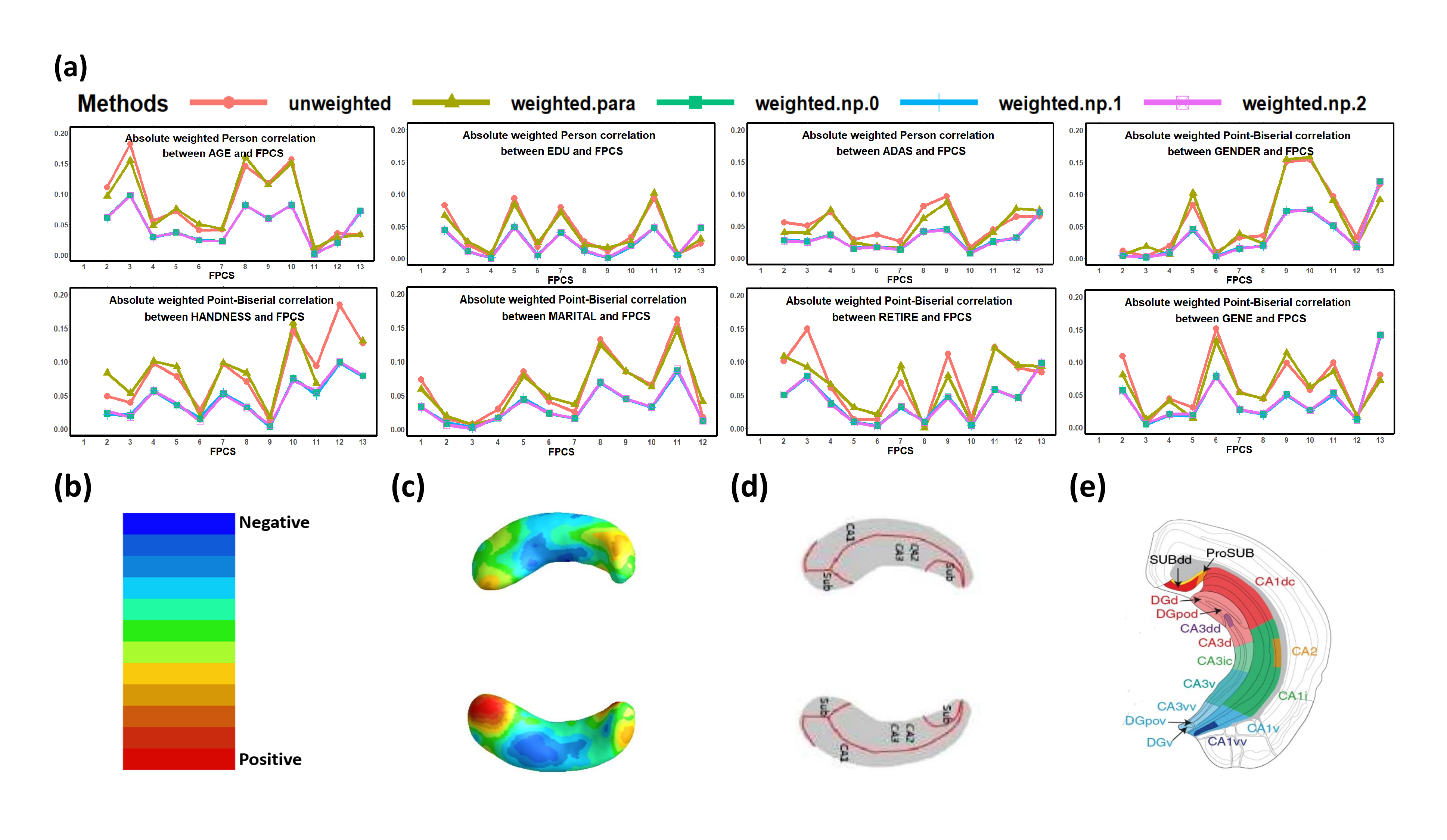}

\caption{ADNI data analysis results: (a) The covariate balancing comparison between the first thirteen FPCS of the image and each covariate when using equal weights (unweighted), parametric weights (weighted.para), and nonparametric weights with $\rho_0=0.1/N$ (weighted.np.0), $\rho_1=1/N$ (weighted.np.1), and $\rho_2=0.01/N$ (weighted.np.2). The weighted absolute Pearson and Point-Biserial correlation is calculated for continuous and categorical covariates, respectively.
(b) The color bar illustration. (c) The estimated coefficient function $\hat{\beta}_{\text{FIPW.np.0}}(\cdot)$. (d) The hippocampal subfields from \citet{2018_Kong}. (e) Newly found refined organizational hippocampal subfields by \citet{bienkowski2018integration}.}
\label{fig:plot_est_balancing_all70}
\end{figure}

Specifically, we applied FPCA to the treatment with the top 12 FPCS ($K_n$=12, PVE=70\%) selected to summarize the image data.
To investigate potential correlations between the functional treatment and each clinical measurement, we calculated the absolute Pearson correlation between each of the top 13 FPCs and each continuous covariate, as well as the absolute Point-Bserial correlation for each categorical covariate.
We reported the weighted absolute Pearson and Point-Bserial correlation to assess the performance of the estimated weights in improving covariate balance.
We calculated weights both parametrically, and nonparametrically using three different values of the tuning parameter: $\rho_0=0.1/N$ (default), $\rho_1=1/N$ (more tolerance of imbalance), and $\rho_2=0.01/N$ (lower tolerance). 
The results are denoted as unweighted, weighted.para, weighted.np.0, weighted.np.1, and weighted.np.2 in \autoref{fig:plot_est_balancing_all70} (a).
As shown in all eight plots, the treatment is correlated with all covariates. 
The usage of weights is able to reduce the correlation over the top 12 FPCs in general, but it is obvious to see that the nonparametrically estimated weights perform better in improving covariate balance conditions.
In addition, nonparametric weighting performs pretty robustly to different $\rho$'s. 
For age, handedness, and retirement, nonparametric weighting effectively lowers down the correlation value of the first FPC and keeps the correlation of the second to twelfth FPCs the lowest. 
In contrast, weights.para reduces the imbalance less, possibly due to misspecified Gaussian distributions when calculating the weights parametrically.

We evaluated the causal effect using all proposed estimators. 
When using three different tuning parameters in FIPW.np estimation, the results are pretty close to each other. 
The FIPW.para also gives similar estimations.
However, the regression adjustment approach yields different estimators from the FIPW approach and the double robust approach performs between them.
This might be because they fit FAFT twice and thus rely more on the imputed outcomes.
Since the censoring rate is pretty high in the study, the resulting estimators are influenced, as shown in the simulation study.
We included all causal estimators in Section 3 of the supplementary material and presented the FIPW.np.0 (FIPW.np using $\rho_0=0.1/N$) estimator in \autoref{fig:plot_est_balancing_all70} (b) and (c).
It can be clearly seen that the subfield of CA1 on both hippocampus has a negative effect on the survival time, indicating that the thicker these areas on the hippocampus are, the shorter the time is to convert to AD. 
Compared to the work of \citet{2018_Kong} and \citet{2021_Yang}, our work identified more precise and refined subregions.
Specifically, the previous works focused on the conditional effect on general subfields as shown in panel (d).
As shown in panel (e), our work examined the causal effect and gave more precise subregions, which is in line with the findings in the medical research \citep{bienkowski2018integration}.
The reason for such a causal effect lies in the accumulation of neurofibrillary tangles (NFTs) first in the CA1 area and then gradually in the subiculum, CA2, CA3, and DG \citep{rao2022hippocampus}, which might be considered as potential targets for early intervention cure.

\section{Discussion}
\label{sec:discussion}
The main contribution of this paper is the development of the FAFT model and a causal inference framework for observational studies involving time-to-event data and infinite-dimensional functional treatments. 
Due to the lack of a direct functional survival model, we propose a FAFT model to incorporate infinite-dimensional imaging data for a survival outcome. 
Three causal inference approaches were developed to adjust confounding effects and result in causal estimators.
The results of the simulation study indicated the appealing numerical performance of the proposed methods in balancing confounding variables, estimating causal effects, and predicting survival outcomes, and demonstrated the robustness with respect to different right censoring rates.
In addition, we successfully identified more precise subregions of the hippocampus brain area and made the first endeavor to study its causal effect on conversion time from MCI to AD. 

The proposed methods may be generalized to other applications. 
For example, it is almost directly applicable to handling multidimensional continuous treatments (e.g., \cite{kong2019multi}), multidimensional functional treatments, and functional/categorical outcomes.
With slight modifications, the proposed methods can be used to study the interval-censored data and other survival models such as the generalized transformation model.
We list several directions for future work.
The proposed weights are only capable of controlling the correlation between treatments and covariates.
In order to improve the stability of weights and the performance of weighted estimators, the weights could be constructed to control the balancing error of the treatments and any outcome function on confounders. 
A possible solution could be to estimate weights via the kernel method based on a reproducing kernel Hilbert space \citep{wang2021estimation}
Another direction is to incorporate a longitudinal assessments of the functional treatment, such as multiple MRI images over 5 years for one subject.
In this situation, a shared confounding structure is needed in order to emphasize the identifiability and estimation of causal effects.  \citep{kong2022identifiability}.
Thus, it is possible to generalize our method by including the latent confounding structure for a better understanding of disease progression and higher accuracy of survival prediction.

\bibliographystyle{asa}
\bibliography{33_ref}

\begin{thebibliography}{27}
\newcommand{\enquote}[1]{``#1''}
\expandafter\ifx\csname natexlab\endcsname\relax\def\natexlab#1{#1}\fi

\bibitem[{{Alzheimer's Association}(2022)}]{AD2022}
{Alzheimer's Association} (2022), \enquote{2022 Alzheimer's disease facts and
  figures,} \textit{Alzheimer's \& Dementia}, 18, 700--789.

\bibitem[{Ashleigh et~al.(2023)Ashleigh, Swerdlow, and Beal}]{ashleigh2023role}
Ashleigh, T., Swerdlow, R.~H., and Beal, M.~F. (2023), \enquote{The role of
  mitochondrial dysfunction in Alzheimer's disease pathogenesis,}
  \textit{Alzheimer's \& Dementia}, 19, 333--342.

\bibitem[{Bienkowski et~al.(2018)Bienkowski, Bowman, Song, Gou, Ard, Cotter,
  Zhu, Benavidez, Yamashita, Abu-Jaber, et~al.}]{bienkowski2018integration}
Bienkowski, M.~S., Bowman, I., Song, M.~Y., Gou, L., Ard, T., Cotter, K., Zhu,
  M., Benavidez, N.~L., Yamashita, S., Abu-Jaber, J., et~al. (2018),
  \enquote{Integration of gene expression and brain-wide connectivity reveals
  the multiscale organization of mouse hippocampal networks,} \textit{Nature
  neuroscience}, 21, 1628--1643.

\bibitem[{Cao and Yu(2023)}]{cao2023adjusting}
Cao, Y. and Yu, J. (2023), \enquote{Adjusting for unmeasured confounding in
  survival causal effect using validation data,} \textit{Computational
  Statistics \& Data Analysis}, 180, 107660.

\bibitem[{Corder and Yang(2022)}]{corder2022utilizing}
Corder, N. and Yang, S. (2022), \enquote{Utilizing stratified generalized
  propensity score matching to approximate blocked randomized designs with
  multiple treatment levels,} \textit{Journal of Biopharmaceutical Statistics},
  32, 373--399.

\bibitem[{Cui et~al.(2021)Cui, Michael, Tanser, and
  Tchetgen~Tchetgen}]{cui2021instrumental}
Cui, Y., Michael, H., Tanser, F., and Tchetgen~Tchetgen, E. (2021),
  \enquote{{Instrumental variable estimation of the marginal structural Cox
  model for time-varying treatments},} \textit{Biometrika}, asab062.

\bibitem[{{HHS, CDC, and NSHS}(2022)}]{CDC2022}
{HHS, CDC, and NSHS} (2022), \enquote{U.S. Department of Health and Human
  Services, Centers for Disease Control and Prevention, National Center for
  Health Statistics. CDC WONDER online database: About Underlying Cause of
  Death, 1999-2019,} CDC WONDER online database: About Underlying Cause of
  Death, 1999-2019. Available at: \url{https://wonder.cdc.gov/ucd-icd10.html}.

\bibitem[{Jiao et~al.(2021)Jiao, Frostig, and Ombao}]{jiao2021filtrated}
Jiao, S., Frostig, R.~D., and Ombao, H. (2021), \enquote{Filtrated Common
  Functional Principal Components for Multivariate Functional data,}
  \textit{arXiv preprint arXiv:2106.01104}.

\bibitem[{Jin et~al.(2006)Jin, Lin, and Ying}]{jin2006least}
Jin, Z., Lin, D., and Ying, Z. (2006), \enquote{On least-squares regression
  with censored data,} \textit{Biometrika}, 93, 147--161.

\bibitem[{Kong et~al.(2018)Kong, Ibrahim, Lee, and Zhu}]{2018_Kong}
Kong, D., Ibrahim, J.~G., Lee, E., and Zhu, H. (2018), \enquote{FLCRM:
  Functional linear cox regression model,} \textit{Biometrics}, 74, 109—117.

\bibitem[{Kong et~al.(2019)Kong, Yang, and Wang}]{kong2019multi}
Kong, D., Yang, S., and Wang, L. (2019), \enquote{Multi-cause causal inference
  with unmeasured confounding and binary outcome,} \textit{arXiv preprint
  arXiv:1907.13323}.

\bibitem[{Kong et~al.(2022)Kong, Yang, and Wang}]{kong2022identifiability}
Kong, D., Yang, S., and Wang, L. (2022), \enquote{Identifiability of causal
  effects with multiple causes and a binary outcome,} \textit{Biometrika}, 109,
  265--272.

\bibitem[{Martinussen(2022)}]{2022_Marti}
Martinussen, T. (2022), \enquote{Causality and the Cox Regression Model,}
  \textit{Annual Review of Statistics and Its Application}, 9, 249--259.

\bibitem[{Potashman et~al.(2023)Potashman, Pang, Tahir, Shahraz, Dichter,
  Perneczky, and Nolte}]{potashman2023psychometric}
Potashman, M., Pang, M., Tahir, M., Shahraz, S., Dichter, S., Perneczky, R.,
  and Nolte, S. (2023), \enquote{Psychometric properties of the Alzheimer’s
  Disease Cooperative Study--Activities of Daily Living for Mild Cognitive
  Impairment (ADCS-MCI-ADL) scale: a post hoc analysis of the ADCS ADC-008
  trial,} \textit{BMC geriatrics}, 23, 124.

\bibitem[{Rao et~al.(2022)Rao, Ganaraja, Murlimanju, Joy, Krishnamurthy, and
  Agrawal}]{rao2022hippocampus}
Rao, Y.~L., Ganaraja, B., Murlimanju, B., Joy, T., Krishnamurthy, A., and
  Agrawal, A. (2022), \enquote{Hippocampus and its involvement in Alzheimer’s
  disease: a review,} \textit{3 Biotech}, 12, 1--10.

\bibitem[{Ritov(1990)}]{1990_Ritov}
Ritov, Y. (1990), \enquote{{Estimation in a Linear Regression Model with
  Censored Data},} \textit{The Annals of Statistics}, 18, 303 -- 328.

\bibitem[{Rubin(1974)}]{rubin1974estimating}
Rubin, D.~B. (1974), \enquote{Estimating causal effects of treatments in
  randomized and nonrandomized studies.} \textit{Journal of educational
  Psychology}, 66, 688.

\bibitem[{Saikia and Barman(2017)}]{saikia2017review}
Saikia, R. and Barman, M.~P. (2017), \enquote{A review on accelerated failure
  time models,} \textit{International Journal of Statistics and Systems}, 12,
  311--322.

\bibitem[{Sang et~al.(2022)Sang, Kong, and Yang}]{sang2022functional}
Sang, P., Kong, D., and Yang, S. (2022), \enquote{Functional principal
  component analysis for longitudinal observations with sampling at random,} .

\bibitem[{Van~der Laan and Robins(2003)}]{van2003unified}
Van~der Laan, M.~J. and Robins, J.~M. (2003), \textit{Unified methods for
  censored longitudinal data and causality}, vol.~5, Springer.

\bibitem[{Wang et~al.(2021)Wang, Wong, Yang, and Chan}]{wang2021estimation}
Wang, J., Wong, R. K.~W., Yang, S., and Chan, K. C.~G. (2021),
  \enquote{Estimation of Partially Conditional Average Treatment Effect by
  Hybrid Kernel-covariate Balancing,} .

\bibitem[{Wang et~al.(2016)Wang, Chiou, and M{\"u}ller}]{wang2016functional}
Wang, J.-L., Chiou, J.-M., and M{\"u}ller, H.-G. (2016), \enquote{Functional
  data analysis,} \textit{Annual Review of Statistics and its application}, 3,
  257--295.

\bibitem[{Warren and Moustafa(2023)}]{warren2023functional}
Warren, S.~L. and Moustafa, A.~A. (2023), \enquote{Functional magnetic
  resonance imaging, deep learning, and Alzheimer's disease: A systematic
  review,} \textit{Journal of Neuroimaging}, 33, 5--18.

\bibitem[{Yang et~al.(2021)Yang, Zhu, Ahn, and Ibrahim}]{2021_Yang}
Yang, H., Zhu, H., Ahn, M., and Ibrahim, J.~G. (2021), \enquote{Weighted
  functional linear Cox regression model,} \textit{Statistical Methods in
  Medical Research}, 30, 1917--1931.

\bibitem[{Yang et~al.(2020)Yang, Pieper, and Cools}]{yang2020semiparametric}
Yang, S., Pieper, K., and Cools, F. (2020), \enquote{Semiparametric estimation
  of structural failure time models in continuous-time processes,}
  \textit{Biometrika}, 107, 123--136.

\bibitem[{Zhang et~al.(2021)Zhang, Xue, and Wang}]{2021_Xiaoke}
Zhang, X., Xue, W., and Wang, Q. (2021), \enquote{Covariate balancing
  functional propensity score for functional treatments in cross-sectional
  observational studies,} \textit{Computational Statistics \& Data Analysis},
  163, 107303.

\bibitem[{Zhao et~al.(2020)Zhao, van Dyk, and Imai}]{zhao2020propensity}
Zhao, S., van Dyk, D.~A., and Imai, K. (2020), \enquote{Propensity score-based
  methods for causal inference in observational studies with non-binary
  treatments,} \textit{Statistical methods in medical research}, 29, 709--727.

\end{thebibliography}

\clearpage
\newpage

\begin{table}[]
\caption{Simulation results for estimation accuracy for the functional causal curve $\beta(\cdot)$ under two scenarios with three censoring rates when the sample size is $N=400$.}
\vspace{3mm}
\label{table:simluation.beta}
\renewcommand\arraystretch{2}
\begin{adjustbox}{width=\columnwidth,center}
\begin{tabular}{lcccclcccclcccc}
\hline
          & \multicolumn{14}{c}{Different right censoring rates}                                                                               \\ \cline{2-15} 
          & \multicolumn{4}{c}{20\%}              &  & \multicolumn{4}{c}{40\%}                 &  & \multicolumn{4}{c}{60\%}                  \\
          & RMSE  & AISE (SE)     & MISE  & ISB   &  & RMSE  & AISE (SE)      & MISE   & ISB    &  & RMSE  & AISE (SE)       & MISE   & ISB    \\ \hline
\multicolumn{15}{l}{Scenario 1}                                                                                                                \\ \hline
Naive     & 0.18  & 0.50(0.036)   & 0.50  & 0.49  &  & 0.18  & 0.50(0.040)    & 0.50   & 0.49   &  & 0.18  & 0.50(0.056)     & 0.50   & 0.49   \\
RegAdj    & 0.00  & 0.00(0.002)   & 0.00  & 0.00  &  & 0.00  & 0.00(0.002)    & 0.00   & 0.00   &  & 0.00  & 0.00(0.003)     & 0.00   & 0.00   \\
FIPW-para & 0.02  & 0.23(0.564)   & 0.12  & 0.06  &  & 0.01  & 0.22(0.590)    & 0.12   & 0.03   &  & 0.01  & 0.30(0.645)     & 0.15   & 0.02   \\
FIPW-np   & 0.02  & 0.09(0.065)   & 0.08  & 0.05  &  & 0.03  & 0.13(0.077)    & 0.11   & 0.08   &  & 0.04  & 0.16(0.091)     & 0.15   & 0.10   \\
DR.para   & 0.00  & 0.00(0.004)   & 0.00  & 0.00  &  & 0.00  & 0.00(0.004)    & 0.00   & 0.00   &  & 0.00  & 0.01(0.004)     & 0.00   & 0.00   \\
DR.np     & 0.00  & 0.00(0.002)   & 0.00  & 0.00  &  & 0.00  & 0.00(0.002)    & 0.00   & 0.00   &  & 0.00  & 0.00(0.003)     & 0.00   & 0.00   \\ \hline
\multicolumn{15}{l}{Scenario 2}                                                                                                                \\ \hline
Naive     & 28.90 & 83.15(30.207) & 77.74 & 79.19 &  & 49.37 & 142.67(57.488) & 131.52 & 135.29 &  & 92.51 & 268.10(115.493) & 239.73 & 253.50 \\
RegAdj    & 4.38  & 14.74(9.766)  & 12.79 & 12.01 &  & 9.94  & 31.82(19.470)  & 26.84  & 27.25  &  & 21.35 & 67.11(42.056)   & 55.76  & 58.52  \\
FIPW-para & 2.63  & 7.92(3.035)   & 7.30  & 7.20  &  & 3.09  & 9.49(4.770)    & 8.44   & 8.47   &  & 4.27  & 13.64(8.566)    & 11.72  & 11.70  \\
FIPW-np   & 1.18  & 3.51(1.809)   & 3.11  & 3.22  &  & 1.32  & 3.96(2.199)    & 3.41   & 3.61   &  & 1.30  & 4.11(2.933)     & 3.27   & 3.55   \\
DR.para   & 1.64  & 6.40(4.824)   & 5.28  & 4.49  &  & 5.72  & 19.53(13.151)  & 16.64  & 15.68  &  & 14.44 & 48.66(34.866)   & 39.58  & 39.57  \\
DR.np     & 2.09  & 6.90(5.405)   & 5.70  & 5.71  &  & 6.58  & 20.93(13.842)  & 17.66  & 18.02  &  & 16.14 & 51.58(34.104)   & 42.33  & 44.23  \\ \hline
\end{tabular}
\end{adjustbox}
\end{table}

\begin{table}[]
\caption{Simulation results for prediction accuracy in terms of Root-MSE for the survival outcome $Y=\log T$ under two scenarios, both in-sample and out-sample, and with three censoring rates when sample size being $N=400$.}
\vspace{3mm}
\label{table:simluation.prediction}
\renewcommand\arraystretch{1.5}
\begin{adjustbox}{width=\columnwidth,center}
\begin{tabular}{llcccccccccccccc}
\hline
                                                                       &           & \multicolumn{14}{c}{Different right censoring rates}                                                                                                                            \\ \cline{3-16} 
                                                                       &           & \multicolumn{4}{c}{20\%}                  & \multicolumn{1}{l}{} & \multicolumn{4}{c}{40\%}                  & \multicolumn{1}{l}{} & \multicolumn{4}{c}{60\%}                  \\ \cline{3-16} 
\multicolumn{2}{l}{}                                                               & Mean  & $q_{.25}$ & $q_{.50}$ & $q_{.75}$ & \multicolumn{1}{l}{} & Mean  & $q_{.25}$ & $q_{.50}$ & $q_{.75}$ & \multicolumn{1}{l}{} & Mean  & $q_{.25}$ & $q_{.50}$ & $q_{.75}$ \\ \hline
\multicolumn{16}{l}{Scenario 1}                                                                                                                                                                                                                                      \\ \hline
\multirow{2}{*}{\begin{tabular}[c]{@{}l@{}}In \\ sample\end{tabular}}  & Naive     & 2.07  & 2.00      & 2.07      & 2.13      &                      & 2.06  & 2.00      & 2.06      & 2.13      &                      & 2.07  & 1.98      & 2.06      & 2.15      \\
                                                                       & RegAdj    & 0.52  & 0.50      & 0.51      & 0.53      &                      & 0.52  & 0.50      & 0.52      & 0.54      &                      & 0.53  & 0.51      & 0.52      & 0.54      \\
                                                                       & FIPW-para & 1.35  & 0.92      & 1.20      & 1.59      &                      & 1.45  & 1.02      & 1.26      & 1.72      &                      & 1.96  & 1.32      & 1.63      & 2.23      \\
                                                                       & FIPW-np   & 1.03  & 0.86      & 1.03      & 1.18      &                      & 1.17  & 0.99      & 1.17      & 1.33      &                      & 1.37  & 1.17      & 1.37      & 1.54      \\
                                                                       & DR.para   & 0.52  & 0.50      & 0.52      & 0.54      &                      & 0.53  & 0.50      & 0.52      & 0.54      &                      & 0.53  & 0.51      & 0.53      & 0.55      \\
                                                                       & DR.np     & 0.52  & 0.50      & 0.52      & 0.53      &                      & 0.52  & 0.50      & 0.52      & 0.54      &                      & 0.53  & 0.51      & 0.52      & 0.55      \\ \hline
\multirow{2}{*}{\begin{tabular}[c]{@{}l@{}}Out \\ sample\end{tabular}} & Naive     & 2.04  & 1.92      & 2.04      & 2.15      &                      & 2.04  & 1.91      & 2.04      & 2.16      &                      & 2.05  & 1.91      & 2.03      & 2.17      \\
                                                                       & RegAdj    & 0.53  & 0.49      & 0.52      & 0.56      &                      & 0.53  & 0.49      & 0.52      & 0.56      &                      & 0.53  & 0.50      & 0.53      & 0.57      \\
                                                                       & FIPW-para & 1.36  & 0.94      & 1.23      & 1.61      &                      & 1.47  & 1.03      & 1.29      & 1.76      &                      & 1.97  & 1.31      & 1.66      & 2.28      \\
                                                                       & FIPW-np   & 1.05  & 0.87      & 1.03      & 1.20      &                      & 1.18  & 1.01      & 1.18      & 1.34      &                      & 1.38  & 1.17      & 1.36      & 1.56      \\
                                                                       & DR.para   & 0.53  & 0.49      & 0.52      & 0.56      &                      & 0.53  & 0.50      & 0.53      & 0.57      &                      & 0.54  & 0.50      & 0.53      & 0.57      \\
                                                                       & DR.np     & 0.53  & 0.49      & 0.52      & 0.56      &                      & 0.53  & 0.50      & 0.52      & 0.56      &                      & 0.54  & 0.50      & 0.53      & 0.57      \\ \hline
\multicolumn{16}{l}{Scenario 2}                                                                                                                                                                                                                                      \\ \hline
\multirow{2}{*}{\begin{tabular}[c]{@{}l@{}}In \\ sample\end{tabular}}  & Naive     & 25.56 & 21.98     & 25.09     & 28.27     &                      & 34.32 & 28.97     & 33.29     & 38.02     &                      & 51.35 & 42.90     & 49.05     & 58.01     \\
                                                                       & RegAdj    & 10.28 & 7.98      & 10.02     & 12.11     &                      & 17.82 & 13.82     & 17.02     & 20.46     &                      & 33.73 & 27.12     & 31.51     & 38.44     \\
                                                                       & FIPW-para & 7.93  & 6.85      & 7.73      & 8.83      &                      & 8.71  & 7.18      & 8.43      & 9.85      &                      & 10.86 & 8.51      & 10.36     & 12.65     \\
                                                                       & FIPW-np   & 5.24  & 4.44      & 5.07      & 5.87      &                      & 5.55  & 4.58      & 5.35      & 6.25      &                      & 5.52  & 4.24      & 5.20      & 6.50      \\
                                                                       & DR.para   & 6.73  & 5.10      & 6.39      & 8.06      &                      & 14.23 & 10.88     & 13.73     & 16.56     &                      & 27.99 & 22.52     & 26.16     & 31.96     \\
                                                                       & DR.np     & 7.06  & 5.20      & 6.80      & 8.51      &                      & 14.27 & 10.99     & 13.49     & 16.54     &                      & 27.95 & 22.44     & 26.59     & 31.88     \\ \hline
\multirow{2}{*}{\begin{tabular}[c]{@{}l@{}}Out \\ sample\end{tabular}} & Naive     & 25.35 & 22.08     & 24.72     & 28.35     &                      & 34.11 & 28.93     & 33.00     & 38.46     &                      & 51.15 & 42.37     & 49.32     & 57.07     \\
                                                                       & RegAdj    & 10.14 & 7.87      & 9.81      & 11.85     &                      & 17.73 & 13.95     & 16.96     & 20.37     &                      & 33.66 & 27.18     & 31.98     & 38.45     \\
                                                                       & FIPW-para & 7.84  & 6.82      & 7.66      & 8.71      &                      & 8.63  & 7.10      & 8.35      & 9.78      &                      & 10.82 & 8.37      & 10.32     & 12.62     \\
                                                                       & FIPW-np   & 5.19  & 4.31      & 4.99      & 5.89      &                      & 5.50  & 4.53      & 5.30      & 6.28      &                      & 5.48  & 4.15      & 5.13      & 6.44      \\
                                                                       & DR.para   & 6.66  & 4.95      & 6.34      & 7.96      &                      & 14.17 & 11.01     & 13.47     & 16.52     &                      & 27.94 & 22.13     & 26.41     & 31.79     \\
                                                                       & DR.np     & 6.97  & 5.14      & 6.76      & 8.48      &                      & 14.20 & 11.03     & 13.68     & 16.58     &                      & 27.89 & 22.18     & 26.44     & 31.91     \\ \hline
\end{tabular}
\end{adjustbox}
\end{table}

\begin{table}[]
\caption{Simulation results of prediction accuracy in terms of Root-MSE for the survival outcome $Y=\log T$ under two scenarios, both in-sample and out-sample, and with three censoring rates when sample size being $N=200$.}
\vspace{3mm}
\label{table:simluation.prediction.N200}
\renewcommand\arraystretch{1.5}
\begin{adjustbox}{width=\columnwidth,center}
\begin{tabular}{llllllllllllllll}
\hline
                                                                       &           & \multicolumn{14}{c}{Different right censoring rates}                                                                                                                                                                                                                                                                                                                                 \\ \cline{3-16} 
                                                                       &           & \multicolumn{4}{c}{20\%}                                                                                                 &  & \multicolumn{4}{c}{40\%}                                                                                                 &  & \multicolumn{4}{c}{60\%}                                                                                                 \\ \cline{3-16} 
\multicolumn{2}{l}{}                                                               & \multicolumn{1}{c}{Mean} & \multicolumn{1}{c}{$q_{.25}$} & \multicolumn{1}{c}{$q_{.50}$} & \multicolumn{1}{c}{$q_{.75}$} &  & \multicolumn{1}{c}{Mean} & \multicolumn{1}{c}{$q_{.25}$} & \multicolumn{1}{c}{$q_{.50}$} & \multicolumn{1}{c}{$q_{.75}$} &  & \multicolumn{1}{c}{Mean} & \multicolumn{1}{c}{$q_{.25}$} & \multicolumn{1}{c}{$q_{.50}$} & \multicolumn{1}{c}{$q_{.75}$} \\ \hline
\multicolumn{16}{l}{Scenario 1}                                                                                                                                                                                                                                                                                                                                                                                                                                           \\ \hline
\multirow{2}{*}{\begin{tabular}[c]{@{}l@{}}In \\ sample\end{tabular}}  & Naive     & 2.08                     & 1.98                          & 2.06                          & 2.17                          &  & 2.08                     & 1.97                          & 2.07                          & 2.18                          &  & 2.09                     & 1.97                          & 2.07                          & 2.21                          \\
                                                                       & RegAdj    & 0.53                     & 0.50                          & 0.53                          & 0.56                          &  & 0.54                     & 0.50                          & 0.53                          & 0.57                          &  & 0.55                     & 0.51                          & 0.55                          & 0.59                          \\
                                                                       & FIPW.para & 1.35                     & 0.94                          & 1.22                          & 1.53                          &  & 1.48                     & 1.02                          & 1.28                          & 1.74                          &  & 1.92                     & 1.29                          & 1.66                          & 2.20                          \\
                                                                       & FIPW.np   & 1.25                     & 1.01                          & 1.25                          & 1.46                          &  & 1.41                     & 1.14                          & 1.41                          & 1.67                          &  & 1.67                     & 1.36                          & 1.63                          & 1.94                          \\
                                                                       & DR.para   & 0.54                     & 0.51                          & 0.53                          & 0.57                          &  & 0.55                     & 0.51                          & 0.54                          & 0.58                          &  & 0.56                     & 0.52                          & 0.55                          & 0.59                          \\
                                                                       & DR.np     & 0.54                     & 0.50                          & 0.53                          & 0.57                          &  & 0.54                     & 0.51                          & 0.54                          & 0.58                          &  & 0.56                     & 0.52                          & 0.55                          & 0.59                          \\ \hline
\multirow{2}{*}{\begin{tabular}[c]{@{}l@{}}Out \\ sample\end{tabular}} & Naive     & 2.03                     & 1.85                          & 2.03                          & 2.20                          &  & 2.03                     & 1.86                          & 2.03                          & 2.21                          &  & 2.04                     & 1.84                          & 2.03                          & 2.23                          \\
                                                                       & RegAdj    & 0.55                     & 0.49                          & 0.54                          & 0.59                          &  & 0.56                     & 0.50                          & 0.54                          & 0.60                          &  & 0.57                     & 0.51                          & 0.56                          & 0.62                          \\
                                                                       & FIPW.para & 1.37                     & 0.94                          & 1.24                          & 1.59                          &  & 1.51                     & 1.02                          & 1.33                          & 1.82                          &  & 1.94                     & 1.31                          & 1.66                          & 2.27                          \\
                                                                       & FIPW.np   & 1.27                     & 1.02                          & 1.24                          & 1.51                          &  & 1.44                     & 1.14                          & 1.42                          & 1.70                          &  & 1.68                     & 1.33                          & 1.61                          & 1.96                          \\
                                                                       & DR.para   & 0.56                     & 0.50                          & 0.55                          & 0.60                          &  & 0.56                     & 0.51                          & 0.55                          & 0.61                          &  & 0.57                     & 0.51                          & 0.56                          & 0.62                          \\
                                                                       & DR.np     & 0.55                     & 0.50                          & 0.55                          & 0.60                          &  & 0.56                     & 0.50                          & 0.55                          & 0.61                          &  & 0.57                     & 0.51                          & 0.56                          & 0.62                          \\ \hline
\multicolumn{16}{l}{Scenario 2}                                                                                                                                                                                                                                                                                                                                                                                                                                           \\ \hline
\multirow{2}{*}{\begin{tabular}[c]{@{}l@{}}In \\ sample\end{tabular}}  & Naive     & 25.91                    & 20.79                         & 24.76                         & 29.82                         &  & 34.75                    & 26.92                         & 32.80                         & 41.22                         &  & 53.31                    & 40.10                         & 50.56                         & 63.61                         \\
                                                                       & RegAdj    & 10.59                    & 7.59                          & 9.84                          & 12.82                         &  & 18.10                    & 13.05                         & 16.75                         & 21.40                         &  & 35.31                    & 25.77                         & 33.07                         & 41.68                         \\
                                                                       & FIPW.para & 7.83                     & 6.52                          & 7.58                          & 8.94                          &  & 8.58                     & 6.75                          & 8.31                          & 9.89                          &  & 10.76                    & 7.54                          & 10.01                         & 12.87                         \\
                                                                       & FIPW.np   & 4.71                     & 3.46                          & 4.42                          & 5.77                          &  & 4.98                     & 3.69                          & 4.65                          & 6.09                          &  & 4.81                     & 3.05                          & 4.25                          & 6.08                          \\
                                                                       & DR.para   & 7.21                     & 4.94                          & 6.64                          & 8.74                          &  & 14.65                    & 10.56                         & 13.49                         & 17.37                         &  & 29.78                    & 21.35                         & 27.39                         & 35.37                         \\
                                                                       & DR.np     & 7.38                     & 4.93                          & 6.76                          & 9.08                          &  & 14.63                    & 10.38                         & 13.51                         & 17.37                         &  & 29.82                    & 21.45                         & 27.68                         & 35.51                         \\ \hline
\multirow{2}{*}{\begin{tabular}[c]{@{}l@{}}Out \\ sample\end{tabular}} & Naive     & 25.46                    & 20.38                         & 24.57                         & 29.29                         &  & 34.36                    & 26.60                         & 33.00                         & 40.44                         &  & 53.01                    & 39.86                         & 50.25                         & 62.31                         \\
                                                                       & RegAdj    & 10.35                    & 7.33                          & 9.75                          & 12.48                         &  & 17.99                    & 13.00                         & 16.57                         & 21.56                         &  & 35.34                    & 25.37                         & 32.94                         & 42.39                         \\
                                                                       & FIPW.para & 7.66                     & 6.37                          & 7.36                          & 8.70                          &  & 8.44                     & 6.72                          & 8.05                          & 9.73                          &  & 10.67                    & 7.65                          & 9.76                          & 12.49                         \\
                                                                       & FIPW.np   & 4.60                     & 3.39                          & 4.34                          & 5.63                          &  & 4.87                     & 3.59                          & 4.53                          & 5.85                          &  & 4.71                     & 2.97                          & 4.16                          & 5.88                          \\
                                                                       & DR.para   & 7.12                     & 4.86                          & 6.63                          & 8.68                          &  & 14.61                    & 10.68                         & 13.35                         & 17.47                         &  & 29.79                    & 21.44                         & 27.85                         & 35.79                         \\
                                                                       & DR.np     & 7.22                     & 4.92                          & 6.74                          & 8.99                          &  & 14.53                    & 10.29                         & 13.34                         & 17.50                         &  & 29.79                    & 21.27                         & 27.78                         & 35.66                         \\ \hline
\end{tabular}
\end{adjustbox}
\end{table}

\begin{table}[]
\caption{Summary statistics for clinical measures after categorizing 373 MRI individuals into four groups according to their observed survival outcome, either the progression time from MRI to AD or the right censored time. The sample mean and standard error are reported for continuous variables. Sample frequency and proportion are reported for categorical variables.}
\vspace{3mm}
\label{table:realdata.summary}
\renewcommand\arraystretch{2}
\begin{adjustbox}{width=\columnwidth,center}
\begin{tabular}{@{}llllllllllllll@{}}
\hline
                        &       & \multicolumn{3}{l}{Continuous: mean(se)} & \multicolumn{9}{l}{Categorical: frequency(proportion)}                                                 \\ \cmidrule(l){3-14} 
                        &       &              &             &             &           &           &           &          & allele.1 &           &          & allele.2  &           \\
                        & Total & age          & edu         & adas        & male      & right     & married   & retire   & geno.2   & geno.3    & geno.4   & geno.3    & geno.4    \\ \midrule
$\widetilde{T}\leqslant$ 1 yr       & 57    & 74(7.3)      & 15(3.5)     & 15(4.8)     & 40(70\%)  & 51(89\%)  & 46(81\%)  & 12(21\%) & 0(0\%)   & 52(91\%)  & 5(9\%)   & 22(39\%)  & 35(61\%)  \\
1 yr $<\widetilde{T}\leqslant$ 2 yr & 117   & 75(7.2)      & 16(2.9)     & 12(4.0)     & 69(59\%)  & 107(91\%) & 92(79\%)  & 15(13\%) & 8(7\%)   & 90(77\%)  & 19(16\%) & 47(40\%)  & 70(60\%)  \\
2 yr $<\widetilde{T}\leqslant$ 3 yr & 91    & 75(7.5)      & 16(2.7)     & 10(4.4)     & 58(64\%)  & 86(95\%)  & 73(80\%)  & 24(26\%) & 10(11\%) & 68(75\%)  & 13(14\%) & 40(44\%)  & 51(56\%)  \\
$\widetilde{T}>$ 3 yr              & 108   & 75(7.4)      & 16(3.0)     & 10(3.9)     & 70(65\%)  & 98(91\%)  & 89(82\%)  & 19(18\%) & 8(7\%)   & 90(83\%)  & 10(9\%)  & 60(56\%)  & 48(44\%)  \\ \midrule
                        & 373   & 75(7.3)      & 16(3.0)     & 12(4.5)     & 237(64\%) & 342(92\%) & 300(80\%) & 70(19\%) & 26(7\%)  & 300(80\%) & 47(13\%) & 169(45\%) & 204(55\%) \\ \bottomrule
\end{tabular}
\end{adjustbox}
\end{table}

\end{document}


\maketitle
\begin{abstract}
In this Supplement, Section~\ref{suppsec:proofs} outlines the proofs of propositions mentioned in the main paper. 
Section~\ref{suppsec:algorithms} presents complete algorithms for the functional causal survival framework, including FAFT estimation, regression adjustment approach, FIPW approach, and double robust approach.
The additional simulation results based on a different sample size are incorporated in Section~\ref{suppsec:simulations}.
Section~\ref{suppsec:realData} includes all causal estimators of the hippocampus in the ADNI study, which are estimated under the proposed causal framework.  
\end{abstract}

\vspace{3mm}

\section{Proofs}
\label{suppsec:proofs}
\subsection{Proof of Proposition 1}
\begin{equation*}
\begin{aligned}
\mathbb{E} \Bigl[ \mathrm{w} Y | X \Bigr] 
& = \mathbb{E}_{\mathbf{Z}} \Bigl[  
\mathbb{E}_{x, \mathbf{Z}} \left[ \mathrm{w} Y | X, \mathbf{Z} \right] \Bigr] \\
& = \int \mathbf{Z}\ \mathbb{E}_{x, \mathbf{Z}} \left[ \mathrm{w} Y | X, \mathbf{Z} \right] f(\mathbf{Z}) \mathrm{~d} \mathbf{Z} \\
& = \int \mathbf{Z} \left[\int \frac{f(\mathbf{Z})}{f(\mathbf{Z} | X) } Y f(Y | X, \mathbf{Z}) \mathrm{~d} Y \right] f(\mathbf{Z}) \mathrm{~d} \mathbf{Z} \\
& = \int \mathbf{Z} \int \frac{f(\mathbf{Z})}{f(\mathbf{Z} | X) } \frac{f(Y,\mathbf{Z} | X)}{f(\mathbf{Z})} \times Y \times f(\mathbf{Z}) \mathrm{~d} Y \mathrm{~d} \mathbf{Z} \\
& = \int \int \frac{f(\mathbf{Z})}{f(\mathbf{Z} | X) } \frac{f(Y | \mathbf{Z}, X) f(\mathbf{Z} | X)}{f(\mathbf{Z})} \times  \mathbf{Z} \times Y \times f(\mathbf{Z}) \mathrm{~d} Y \mathrm{~d} \mathbf{Z} \\
& = \int \int Y \times f(Y | \mathbf{Z}, X) \mathrm{~d} Y \times \mathbf{Z} \times f(\mathbf{Z}) \mathrm{~d} \mathbf{Z} \\
& = \int_\mathbf{Z} \mathbb{E}\left[  Y(x) |  \mathbf{Z}=\mathbf{z} \right] \mathrm{~d} f_\mathbf{Z}(\mathbf{z}) \\
& = \int_\mathbf{Z} \mathbb{E}\left[  Y | X = x,  \mathbf{Z} = \mathbf{z} \right] \mathrm{~d} f_\mathbf{Z}(\mathbf{z}) \\
& = \mathbb{E}[Y(x)]
\end{aligned}
\end{equation*}

\subsection{Proof of Proposition 2}
\begin{equation*}
\mathrm{w}
\overset{\triangle}{=} \frac{f(\mathbf{Z})}{f(\mathbf{Z} | X) } 
\approx \frac{f(\mathbf{Z})}{f(\mathbf{Z} | \mathbf{A}) } 
= \frac{f(\mathbf{Z})}{f(\mathbf{A} | \mathbf{Z}) f(\mathbf{Z}) / f(\mathbf{A}) } 
= \frac{f(\mathbf{A})}{f(\mathbf{A} | \mathbf{Z})} \overset{\triangle}{=} w
\end{equation*}

\section{Algorithms}
\label{suppsec:algorithms}
\subsection{Algorithm 1: FAFT estimation}
Equations included in this section correspond to the main paper.

\vspace{5mm}

\noindent
Input: $\left\{\{X_i(s_{1m}), 1 \leqslant m \leqslant M\}, \mathbf{Z}_i, \delta_i, \widetilde{T}_i \ \mid \ i=1,...,n \right\}$
\begin{enumerate}[leftmargin=2.5\parindent]
        \item[Step-1: ] Do FPCA based on pre-determined PVE and return FPCS matrix $\left[\hat{A}_{ik} \right]_{n \times K_n}$ and corresponding eigenfunctions $\{\hat{\phi}_{k}(\cdot): 1 \leqslant k \leqslant K_n \}$. 
	\item[Step-2: ] Evaluate conditional expectations for censored subjects using Equation (10).  
	\item[Step-3: ] Calculate pseudo outcome $Y_i^*$ for each individual based on Equation (9).
	\item[Step-4: ] Given an initial value $\hat{\boldsymbol{\theta}}^{(0)}$, at the $m$-th interation, update $\hat{\boldsymbol{\theta}}^{(m)}_n=L_n\left(\hat{\boldsymbol{\theta}}^{(m-1)}_n\right)$ based on Equation (11). 
	\item[Step-5: ] Repeat Step.4 until predetermined converging criteria is satisfied or maximum interation is reached.
	\item[Step-6: ] Recover $\hat{\beta}(\cdot)$ using estimated $\{ \hat{\beta}_k, \hat{\phi}_{k}(\cdot): 1 \leqslant k \leqslant K_n \}$.
\end{enumerate}
Output: $\left(\hat{\alpha}, \hat{\beta}(\cdot), \hat{\gamma}_1, ..., \hat{\gamma}_p\right)$   
\subsection{Algorithm 2: Regression adjustment approach}
Equations included in this section correspond to the main paper.

\vspace{5mm}

\noindent
Input: $\left\{\{X_i(s_{1m}), 1 \leqslant m \leqslant M\}, \mathbf{Z}_i, \delta_i, \widetilde{T}_i \ \mid \ i=1,...,n \right\}$
\begin{enumerate}[leftmargin=2.5\parindent]
    \item[Step-1: ] Fit FAFT using \hyperref[table:algorithm.faft]{Algorithm 1}.  
    \item[Step-2: ] Construct $n$ adjusted responses based on Equation 3.2 and return $\widehat{\mathbf{{Y}}} = \left( \widehat{Y}_{1}, ..., \widehat{Y}_{n} \right )$ 
    \item[Step-3: ] Fit FAFT model with $\mathbf{\widehat{Y}}$ using \hyperref[table:algorithm.faft]{Algorithm 1}. 
    \item[Step-4: ] Recover $\hat{\beta}_{\text{RegAdj}}(\cdot)$.
\end{enumerate}
Output: $\hat{\beta}_{\text{RegAdj}}(\cdot)$

\subsection{Algorithm 3: FIPW approach}
Equations included in this section correspond to this section, not the main paper.

Ideally, the functional weights are expected to achieve the balance condition for each observed individual, i.e.,
\begin{equation}
\mathbb{E}\left(w_i^{*} \mathbf{A}_{i}^{*} \mathbf{Z}_{i}^{*\top}\right) =0, \quad i=1,...,n.
\label{eq:ipw.balance.condition}
\end{equation}
To satisfy it, $w_i$ can be viewed as a minimizer of the left side of the \autoref{eq:ipw.balance.condition}.
According to \citet{2021_Xiaoke}, the weights can be calculated parametrically or non-parametrically.
The parametric estimation assumes that $f\left(\mathbf{A}^{*}\right) \sim N(0,1)$ and  $f\left(\mathbf{A}^{*} \mid \mathbf{Z}_{i}^{*}\right)  \sim N(\boldsymbol{\xi}^{\top}\mathbf{Z}_{i}^{*}, \boldsymbol{\Sigma}_{\mathbf{A}_{i}^{*}} )$ and unknown parameters $\boldsymbol{\xi}$ and $\Sigma_{A^*}$ can be estimated by solving the following equations:
\begin{equation}
\left\{\begin{array}{l}
n^{-1} \sum_{i=1}^{n}\left(\mathbf{A}_{i}^{*}-\boldsymbol{\xi}^{\top} \mathbf{Z}_{i}^{*}\right)\left(\mathbf{A}_{i}^{*}-\boldsymbol{\xi}^{\top} \mathbf{Z}_{i}^{*}\right)^{\top}=\boldsymbol{\Sigma}_{\mathbf{A}_{i}^{*}} ; \\
n^{-1} \sum_{i=1}^{n} \operatorname{det}(\boldsymbol{\Sigma}_{\mathbf{A}_{i}^{*}})^{1 / 2} \exp \left\{\frac{1}{2}\left(\mathbf{A}_{i}^{*}-\boldsymbol{\xi}^{\top} \mathbf{Z}_{i}^{*}\right)^{\top} \boldsymbol{\Sigma}_{\mathbf{A}_{i}^{*}}^{-1}\left(\mathbf{A}_{i}^{*}-\boldsymbol{\xi}^{\top} \mathbf{Z}_{i}^{*}\right)-\frac{1}{2}\left(\mathbf{A}_{i}^{*}\right)^{\top}\left(\mathbf{A}_{i}^{*}\right)\right\} \mathbf{A}_{i}^{*}\left(\mathbf{Z}_{i}^{*}\right)^{\top}=\mathbf{0} .
\end{array}\right.
\label{eq:weight.par}
\end{equation}
With estimates $\widehat{\boldsymbol{\xi}}$ and $\widehat{\boldsymbol{\Sigma}}_{\mathbf{A}_{i}^{*}}$, it's very straightforward to calculate individual weights
$$
w_i^* = \operatorname{det}(\boldsymbol{\Sigma}_{\mathbf{A}_{i}^{*}})^{1 / 2} \exp \left\{\frac{1}{2}\left(\mathbf{A}_{i}^{*}-\boldsymbol{\xi}^{\top} \mathbf{Z}_{i}^{*}\right)^{\top} \boldsymbol{\Sigma}_{\mathbf{A}_{i}^{*}}^{-1}\left(\mathbf{A}_{i}^{*}-\boldsymbol{\xi}^{\top} \mathbf{Z}_{i}^{*}\right)-\frac{1}{2}\left(\mathbf{A}_{i}^{*}\right)^{\top}\left(\mathbf{A}_{i}^{*}\right)\right\}.
$$

The non-parametric calculation avoids the possible misspecification for the SFPS while sacrificing some computation efficiency. \autoref{eq:ipw.balance.condition}) should be minimized with four restrictions shown as below:
\begin{equation}
\begin{aligned}
E\left(w_{i}^{*}\right) &=1, \quad E\left(w_{i}^{*} \mathrm{~A}_{i}^{*}\mathbf{Z}_{i}^{*\top}\right)=\mathbf{0}, \\
E\left(w_{i}^{*} \mathbf{A}_{i}^{*}\right) &=\mathbf{0}, \quad E\left(w_{i}^{*} \mathbf{Z}_{i}^{*}\right)=\mathbf{0}, \quad i=1, \ldots, n
\end{aligned}
\label{eq:ipw.balance.nonpar}
\end{equation}
Based on the idea of empirical likelihood method \citep{owen2001empirical}, subject to the empirical counterparts of \autoref{eq:ipw.balance.nonpar}, the aim is to maximize $\prod_{i=1}^{n} f_{\left(\mathbf{A}^{*}, \mathbf{Z}^{*}\right)}\left(\mathbf{A}_{i}^{*}, \mathbf{Z}_{i}^{*}\right)$, thus leads to an equivalent optimization:
$$
\min_{\mathbf{w}^{*}} \sum_{i=1}^{n} \log \left(w_{i}^{*}\right),
$$
$$
\text { s.t. } \quad \sum_{i=1}^{n} w_{i}^{*}=n, \quad \sum_{i=1}^{n} w_{i}^{*} \mathbf{A}_{i}^{*}\left(\mathbf{Z}_{i}^{*}\right)^{\top}=\mathbf{0}, \quad  \sum_{i=1}^{n} w_{i}^{*} \mathbf{A}_{i}^{*}=\mathbf{0}, \quad \sum_{i=1}^{n} w_{i}^{*} \mathbf{Z}_{i}^{*}=\mathbf{0}.
$$

Due to the existence of a non-convexity issue, the regularized approach by \citet{fong2018covariate} is adapted to allow an imbalance between $\mathbf{A}_{i}^{*}$ and $\mathbf{Z}_{i}^{*}$ but meanwhile penalizes such imbalance in the objective function

\begin{equation}
\min_{\mathbf{w}^*, \boldsymbol{\Gamma}} 
\sum_{i=1}^{n} \log \left[\sum_{i=1}^n \log \left(w_i^*\right)+\frac{1}{2 \rho}\{\operatorname{vec}(\boldsymbol{\Gamma})\}^{\top}\{\operatorname{vec}(\boldsymbol{\Gamma})\}\right],
\label{eq:weight.nonpar}
\end{equation}
$$
\text { s.t. } \quad \sum_{i=1}^{n} w_{i}^{*}=n, \quad \frac{1}{n} \sum_{i=1}^n w_i^* \mathbf{A}_i^*\left(\mathbf{C}_i^*\right)^{\top}=\boldsymbol{\Gamma}, \quad \sum_{i=1}^{n} w_{i}^{*} \mathbf{A}_{i}^{*}=\mathbf{0}, \quad \sum_{i=1}^{n} w_{i}^{*} \mathbf{Z}_{i}^{*}=\mathbf{0},
$$
where $\rho>0$ is a tuning parameter and $\Gamma \text { is a } L \times p $ matrix allowing for an imperfect balancing condition. Optimization involves the Lagrangian multiplier, profile method, and use of the Fletcher–Goldfarb–Shanno algorithm. The default value of $\rho$ is set to be $0.1/n$ suggested by \cite{fong2018covariate}. More details can be found in \cite{2021_Xiaoke}. 
After getting $\hat{w}_{i}^{*}$ 's, a weighted pseudo-sample created by $\boldsymbol{w} \mathbf{Y}$ will be used to fit FAFT.
The following algorithm summarizes the whole procedure.

\vspace{1cm}

\noindent
Input:  $\left\{\{X_i(s_{1m}), 1 \leqslant m \leqslant M\}, \mathbf{Z}_i, \delta_i, \widetilde{T}_i \ \mid \ i=1,...,n \right\}$
\begin{enumerate}[leftmargin=2.5\parindent]
    \item[Step-1: ] Do FPCA based on pre-determined PVE and return standardized FPCS matrix $\left[\hat{A}^*_{ik} \right]_{n \times K_n}$.
	\item[Step-2: ] Calculate standardized weights $\hat{w}_i$ by solving \autoref{eq:weight.par} parametrically or \autoref{eq:weight.nonpar} nonparametrically
	\item[Step-3: ] Fit FAFT model using \hyperref[table:algorithm.faft]{Algorithm 1} with created pseudo-sample
	$$\left\{\hat{w}_i, \widetilde{T}_i = \min(T_i, C_i), \delta_i, X_i(s), \mathbf{Z}_i | i=1,...,n \right\}.$$
	\item[Step-4: ] Recover $\hat{\beta}_{\text{FIPW}}(\cdot)$.   
\end{enumerate}
Output: $\hat{\beta}_{\text{FIPW}}(\cdot)$

\subsection{Algorithm 4: double robust approach}
Equations included in this section correspond to the main paper.

\vspace{5mm}

\noindent
Input:  $\left\{\{X_i(s_{1m}), 1 \leqslant m \leqslant M\}, \mathbf{Z}_i, \delta_i, \widetilde{T}_i \ \mid \ i=1,...,n \right\}$   
\begin{enumerate}[leftmargin=2.5\parindent]
    \item[Step-1: ] Calculate fitted adjustment responses $\left\{ \hat{Y}_{\text{RegAdj},1}, ..., \hat{Y}_{\text{RegAdj},n} \right \}$ using \hyperref[table:algorithm.regadj]{Algorithm 2}. 
	\item[Step-2: ] Calculate weights $\left\{w_1^*, ..., w_n^*\right \}$ using \hyperref[table:algorithm.fipw]{Algorithm 3}.
	\item[Step-3: ] Construct pseudo outcomes adjusted by weighted residuals $\left\{ \widetilde{Y}_{\text{RegAdj},1}, ..., \widetilde{Y}_{\text{RegAdj},n}\right \}$ based on Equation (8). 
	\item[Step-4: ] Fit FAFT model using \hyperref[table:algorithm.faft]{Algorithm 1}. 
	\item[Step-5: ] Recover $\hat{\beta}_{\text{DR}}(\cdot)$.  
\end{enumerate}
Output: $\hat{\beta}_{\text{DR}}(\cdot)$     
\section{Additional simulation results}
\label{suppsec:simulations}
\begin{figure}[]
\centering
\includegraphics[width=\textwidth]{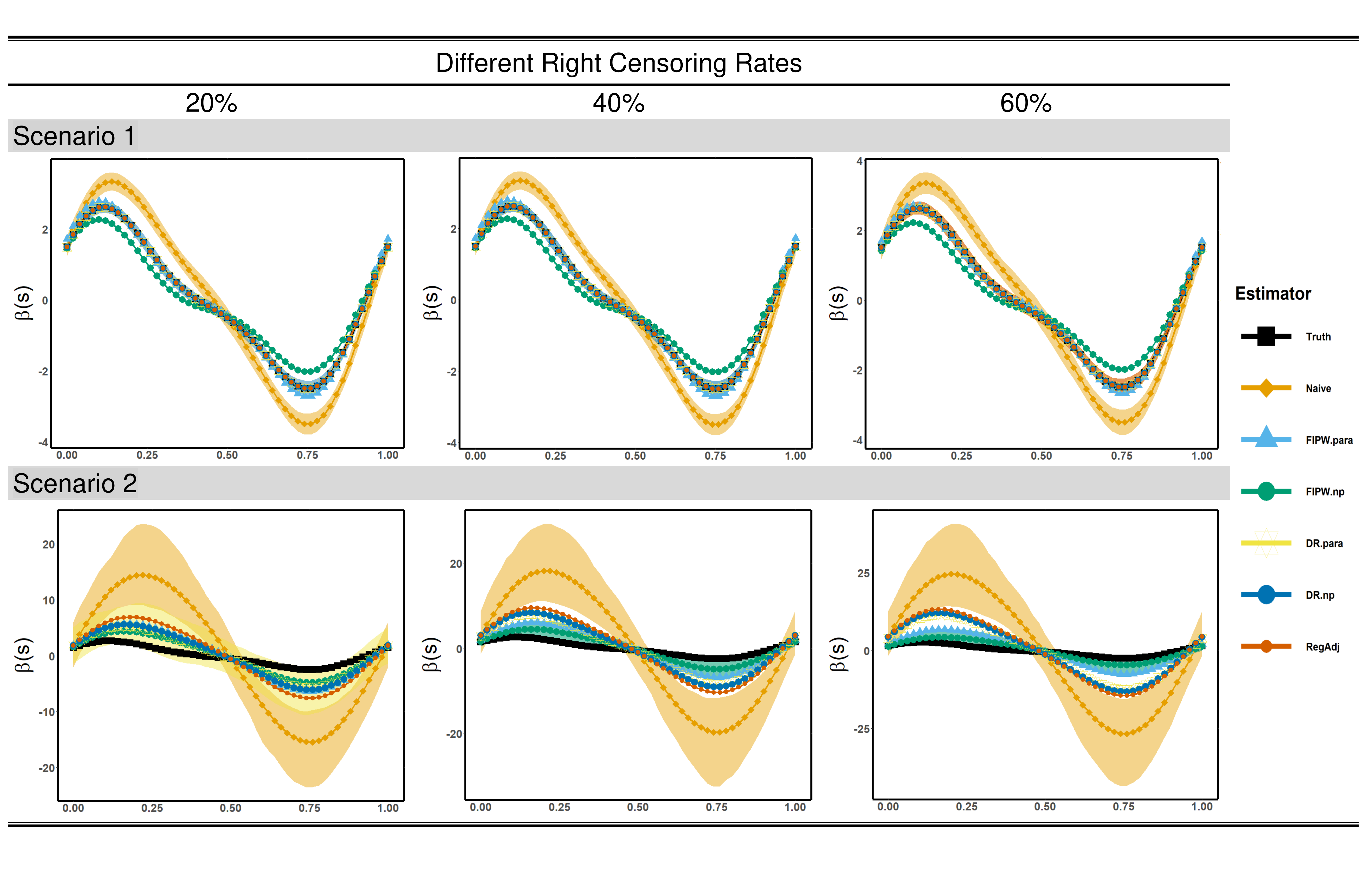}

\caption{Plot of estimated functional causal effect curve under two scenarios with three censoring rates when sample size N=200. Different estimators are represented by line color and point shape with shading area as corresponding CI.}
\label{fig:simulation.beta.N200.plot}
\end{figure}

In this section, we report the finite sample performance of our method based on sample size $N=200$, as censoring rate varies from 20\% to 60\%. 
\autoref{fig:simulation.beta.N200.plot} vitalizes different functional causal estimates.
\autoref{table:simluation.beta.N200} summarizes all results of estimation accuracy for $\beta(\cdot)$.
\autoref{table:simluation.prediction.N200} summarizes all results of causal prediction accuracy for the survival outcome $Y$.

\section{Additional Real Data Results}
\label{suppsec:realData}
\begin{figure}[]
\centering
\includegraphics[width=\textwidth]{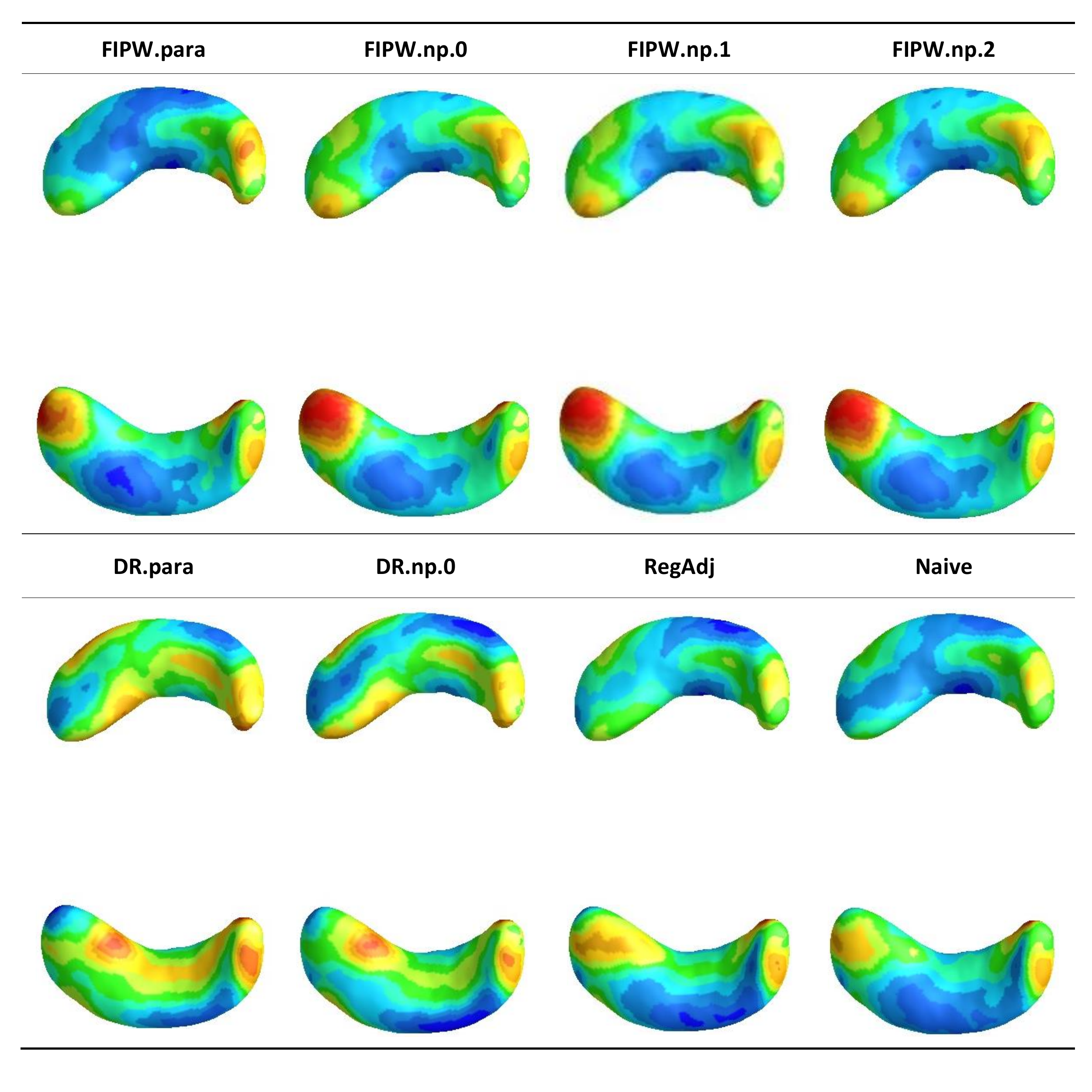}

\caption{ADNI Data Analysis Results Using Different Estimators}
\label{fig:realdata.beta.all70.plot}
\end{figure}

We present all causal estimators in \autoref{fig:realdata.beta.all70.plot}. 
The first row includes four estimators obtained via FIPW approach, and the weights are estimated parametrically, or non-parametrically with three different tuning parameter values.
They are denoted as FIPW.para, FIPW.np.0 ($\rho_0=0.1/n$), FIPW.np.1 ($\rho_1=1/n$), FIPW.np.0 ($\rho_2=0.01/n$), respectively.
In the second row, the double robust approach provides two estimators, DR.para (parametrically estimated weights) and DR.np.0 (non-parametrically estimated weights with $\rho_0=0.1/n$). 
The 'RegADj' represents the estimator using the regression adjustment approach.
For comparison, the naive estimator is also included, which assumes there is no confounding effect.

\backmatter
\bibliographystyle{asa}
\bibliography{33_ref}

\clearpage
\newpage

\begin{table}[]
\caption{Simulation results of estimation accuracy for the functional causal curve $\beta(\cdot)$ under two scenarios with three censoring rates when sample size being $N=200$.}
\label{table:simluation.beta.N200}
\renewcommand\arraystretch{2}
\begin{adjustbox}{width=\columnwidth,center}
\begin{tabular}{@{}lllllllllllllll@{}}
\hline
          & \multicolumn{14}{c}{Different right censoring rates}                                                                               \\ \midrule
          & \multicolumn{4}{c}{20\%}              &  & \multicolumn{4}{c}{40\%}                 &  & \multicolumn{4}{c}{60\%}                  \\ \cmidrule(l){2-15} 
          & RMSE  & AISE (SE)     & MISE  & ISB   &  & RMSE  & AISE (SE)      & MISE   & ISB    &  & RMSE  & AISE (SE)       & MISE   & ISB    \\ \midrule
\multicolumn{15}{l}{Scenario 1}                                                                                                                \\ \midrule
Naive     & 0.18  & 0.51(0.058)   & 0.51  & 0.49  &  & 0.18  & 0.51(0.064)    & 0.51   & 0.49   &  & 0.18  & 0.52(0.086)     & 0.52   & 0.49   \\
RegAdj    & 0.00  & 0.01(0.004)   & 0.00  & 0.00  &  & 0.00  & 0.01(0.004)    & 0.01   & 0.00   &  & 0.00  & 0.01(0.007)     & 0.01   & 0.00   \\
FIPW.para & 0.01  & 0.25(0.340)   & 0.15  & 0.03  &  & 0.01  & 0.28(0.407)    & 0.16   & 0.02   &  & 0.00  & 0.38(0.585)     & 0.23   & 0.01   \\
FIPW.np   & 0.03  & 0.17(0.141)   & 0.14  & 0.07  &  & 0.04  & 0.23(0.165)    & 0.19   & 0.11   &  & 0.05  & 0.29(0.214)     & 0.25   & 0.12   \\
DR.para   & 0.00  & 0.01(0.005)   & 0.01  & 0.00  &  & 0.00  & 0.01(0.005)    & 0.01   & 0.00   &  & 0.00  & 0.01(0.007)     & 0.01   & 0.00   \\
DR.np     & 0.00  & 0.01(0.004)   & 0.01  & 0.00  &  & 0.00  & 0.01(0.005)    & 0.01   & 0.00   &  & 0.00  & 0.01(0.007)     & 0.01   & 0.00   \\ \midrule
\multicolumn{15}{l}{Scenario 2}                                                                                                                \\ \midrule
Naive     & 29.31 & 88.42(46.197) & 77.78 & 80.31 &  & 50.06 & 151.70(85.456) & 132.25 & 137.17 &  & 97.51 & 296.14(171.799) & 254.00 & 267.19 \\
RegAdj    & 4.38  & 17.14(16.232) & 12.93 & 12.00 &  & 9.94  & 35.65(30.804)  & 27.43  & 27.23  &  & 22.63 & 78.25(66.232)   & 60.60  & 62.00  \\
FIPW.para & 2.53  & 7.89(3.944)   & 7.07  & 6.94  &  & 2.98  & 9.69(6.810)    & 8.24   & 8.16   &  & 4.14  & 14.46(14.887)   & 10.96  & 11.34  \\
FIPW.np   & 0.89  & 2.96(2.115)   & 2.33  & 2.44  &  & 1.00  & 3.36(2.555)    & 2.61   & 2.73   &  & 0.91  & 3.49(3.476)     & 2.32   & 2.49   \\
DR.para   & 1.71  & 7.94(8.262)   & 5.76  & 4.68  &  & 5.82  & 22.70(20.494)  & 16.74  & 15.94  &  & 15.73 & 58.93(53.792)   & 44.70  & 43.11  \\
DR.np     & 2.18  & 8.25(9.997)   & 5.86  & 5.98  &  & 6.80  & 24.34(23.151)  & 18.28  & 18.64  &  & 17.96 & 64.10(57.137)   & 48.63  & 49.20  \\ \bottomrule
\end{tabular}
\end{adjustbox}
\end{table}

\begin{table}[]
\caption{Simulation results of prediction accuracy in terms of Root-MSE for the survival outcome $Y=\log T$ under two scenarios, both in-sample and out-sample, and with three censoring rates when sample size being $N=200$.}
\label{table:simluation.prediction.N200}
\renewcommand\arraystretch{1.6}
\begin{adjustbox}{width=\columnwidth,center}
\begin{tabular}{llllllllllllllll}
\hline
                                                                       &           & \multicolumn{14}{c}{Different right censoring rates}                                                                                                                                                                                                                                                                                                                                 \\ \cline{3-16} 
                                                                       &           & \multicolumn{4}{c}{20\%}                                                                                                 &  & \multicolumn{4}{c}{40\%}                                                                                                 &  & \multicolumn{4}{c}{60\%}                                                                                                 \\ \cline{3-16} 
\multicolumn{2}{l}{}                                                               & \multicolumn{1}{c}{Mean} & \multicolumn{1}{c}{$q_{.25}$} & \multicolumn{1}{c}{$q_{.50}$} & \multicolumn{1}{c}{$q_{.75}$} &  & \multicolumn{1}{c}{Mean} & \multicolumn{1}{c}{$q_{.25}$} & \multicolumn{1}{c}{$q_{.50}$} & \multicolumn{1}{c}{$q_{.75}$} &  & \multicolumn{1}{c}{Mean} & \multicolumn{1}{c}{$q_{.25}$} & \multicolumn{1}{c}{$q_{.50}$} & \multicolumn{1}{c}{$q_{.75}$} \\ \hline
\multicolumn{16}{l}{Scenario 1}                                                                                                                                                                                                                                                                                                                                                                                                                                           \\ \hline
\multirow{2}{*}{\begin{tabular}[c]{@{}l@{}}In \\ sample\end{tabular}}  & Naive     & 2.08                     & 1.98                          & 2.06                          & 2.17                          &  & 2.08                     & 1.97                          & 2.07                          & 2.18                          &  & 2.09                     & 1.97                          & 2.07                          & 2.21                          \\
                                                                       & RegAdj    & 0.53                     & 0.50                          & 0.53                          & 0.56                          &  & 0.54                     & 0.50                          & 0.53                          & 0.57                          &  & 0.55                     & 0.51                          & 0.55                          & 0.59                          \\
                                                                       & FIPW.para & 1.35                     & 0.94                          & 1.22                          & 1.53                          &  & 1.48                     & 1.02                          & 1.28                          & 1.74                          &  & 1.92                     & 1.29                          & 1.66                          & 2.20                          \\
                                                                       & FIPW.np   & 1.25                     & 1.01                          & 1.25                          & 1.46                          &  & 1.41                     & 1.14                          & 1.41                          & 1.67                          &  & 1.67                     & 1.36                          & 1.63                          & 1.94                          \\
                                                                       & DR.para   & 0.54                     & 0.51                          & 0.53                          & 0.57                          &  & 0.55                     & 0.51                          & 0.54                          & 0.58                          &  & 0.56                     & 0.52                          & 0.55                          & 0.59                          \\
                                                                       & DR.np     & 0.54                     & 0.50                          & 0.53                          & 0.57                          &  & 0.54                     & 0.51                          & 0.54                          & 0.58                          &  & 0.56                     & 0.52                          & 0.55                          & 0.59                          \\ \hline
\multirow{2}{*}{\begin{tabular}[c]{@{}l@{}}Out \\ sample\end{tabular}} & Naive     & 2.03                     & 1.85                          & 2.03                          & 2.20                          &  & 2.03                     & 1.86                          & 2.03                          & 2.21                          &  & 2.04                     & 1.84                          & 2.03                          & 2.23                          \\
                                                                       & RegAdj    & 0.55                     & 0.49                          & 0.54                          & 0.59                          &  & 0.56                     & 0.50                          & 0.54                          & 0.60                          &  & 0.57                     & 0.51                          & 0.56                          & 0.62                          \\
                                                                       & FIPW.para & 1.37                     & 0.94                          & 1.24                          & 1.59                          &  & 1.51                     & 1.02                          & 1.33                          & 1.82                          &  & 1.94                     & 1.31                          & 1.66                          & 2.27                          \\
                                                                       & FIPW.np   & 1.27                     & 1.02                          & 1.24                          & 1.51                          &  & 1.44                     & 1.14                          & 1.42                          & 1.70                          &  & 1.68                     & 1.33                          & 1.61                          & 1.96                          \\
                                                                       & DR.para   & 0.56                     & 0.50                          & 0.55                          & 0.60                          &  & 0.56                     & 0.51                          & 0.55                          & 0.61                          &  & 0.57                     & 0.51                          & 0.56                          & 0.62                          \\
                                                                       & DR.np     & 0.55                     & 0.50                          & 0.55                          & 0.60                          &  & 0.56                     & 0.50                          & 0.55                          & 0.61                          &  & 0.57                     & 0.51                          & 0.56                          & 0.62                          \\ \hline
\multicolumn{16}{l}{Scenario 2}                                                                                                                                                                                                                                                                                                                                                                                                                                           \\ \hline
\multirow{2}{*}{\begin{tabular}[c]{@{}l@{}}In \\ sample\end{tabular}}  & Naive     & 25.91                    & 20.79                         & 24.76                         & 29.82                         &  & 34.75                    & 26.92                         & 32.80                         & 41.22                         &  & 53.31                    & 40.10                         & 50.56                         & 63.61                         \\
                                                                       & RegAdj    & 10.59                    & 7.59                          & 9.84                          & 12.82                         &  & 18.10                    & 13.05                         & 16.75                         & 21.40                         &  & 35.31                    & 25.77                         & 33.07                         & 41.68                         \\
                                                                       & FIPW.para & 7.83                     & 6.52                          & 7.58                          & 8.94                          &  & 8.58                     & 6.75                          & 8.31                          & 9.89                          &  & 10.76                    & 7.54                          & 10.01                         & 12.87                         \\
                                                                       & FIPW.np   & 4.71                     & 3.46                          & 4.42                          & 5.77                          &  & 4.98                     & 3.69                          & 4.65                          & 6.09                          &  & 4.81                     & 3.05                          & 4.25                          & 6.08                          \\
                                                                       & DR.para   & 7.21                     & 4.94                          & 6.64                          & 8.74                          &  & 14.65                    & 10.56                         & 13.49                         & 17.37                         &  & 29.78                    & 21.35                         & 27.39                         & 35.37                         \\
                                                                       & DR.np     & 7.38                     & 4.93                          & 6.76                          & 9.08                          &  & 14.63                    & 10.38                         & 13.51                         & 17.37                         &  & 29.82                    & 21.45                         & 27.68                         & 35.51                         \\ \hline
\multirow{2}{*}{\begin{tabular}[c]{@{}l@{}}Out \\ sample\end{tabular}} & Naive     & 25.46                    & 20.38                         & 24.57                         & 29.29                         &  & 34.36                    & 26.60                         & 33.00                         & 40.44                         &  & 53.01                    & 39.86                         & 50.25                         & 62.31                         \\
                                                                       & RegAdj    & 10.35                    & 7.33                          & 9.75                          & 12.48                         &  & 17.99                    & 13.00                         & 16.57                         & 21.56                         &  & 35.34                    & 25.37                         & 32.94                         & 42.39                         \\
                                                                       & FIPW.para & 7.66                     & 6.37                          & 7.36                          & 8.70                          &  & 8.44                     & 6.72                          & 8.05                          & 9.73                          &  & 10.67                    & 7.65                          & 9.76                          & 12.49                         \\
                                                                       & FIPW.np   & 4.60                     & 3.39                          & 4.34                          & 5.63                          &  & 4.87                     & 3.59                          & 4.53                          & 5.85                          &  & 4.71                     & 2.97                          & 4.16                          & 5.88                          \\
                                                                       & DR.para   & 7.12                     & 4.86                          & 6.63                          & 8.68                          &  & 14.61                    & 10.68                         & 13.35                         & 17.47                         &  & 29.79                    & 21.44                         & 27.85                         & 35.79                         \\
                                                                       & DR.np     & 7.22                     & 4.92                          & 6.74                          & 8.99                          &  & 14.53                    & 10.29                         & 13.34                         & 17.50                         &  & 29.79                    & 21.27                         & 27.78                         & 35.66                         \\ \hline
\end{tabular}
\end{adjustbox}
\end{table}